# The algorithmic behavioral sink: Modeling the transition to irreversible demographic collapse

Feng Yu

Affiliation: Lawrence Berkeley National Laboratory (LBNL)

*This work is independent research and does not represent the views of LBNL or the U.S. Department of Energy (DOE).*

## Abstract

In 2025, the vast majority of major global economies reported a Total Fertility Rate (TFR) below the replacement level of 2.1. This synchronized collapse has persisted despite extensive pro-natalist financial interventions, revealing a structural limitation in conventional linear demographic models. This paper proposes that the ubiquity of mobile internet and Artificial Intelligence (AI) acts as a synthetic mating disruptor, precipitating a non-linear phase transition in human pair-bonding dynamics. Treating the marriage market as a complex system governed by thermodynamic constraints, we present a Fermi-Dirac selection model—using national income distributions as the input spectrum and fitted with Japanese longitudinal data (2012–2024)—to illustrate how digital "hyper-reality" generates an increasingly prohibitive "Marriage Energy Barrier." Our model quantifies a "Tech Factor" that decouples female mating expectations from local economic realities during initial courtship, a distortion we predict will be severely exacerbated as AI evolves into an emotional substitute. Projections indicate that this substitution effect is accelerating; without intervention, the world post-2040 risks entering an irreversible demographic contraction, as the sudden surge in TFR required for recovery is biologically impossible. Ultimately, this framework explains why traditional financial subsidies fail to reverse falling birth rates and suggests that digital convenience may represent a behavioral "Great Filter" that erodes the imperative for species persistence.

**Keywords:** Artificial Intelligence, Demographic Collapse, Fermi-Dirac Statistics, Hyper-reality, Fertility Rates, Sociophysics

## 1. Introduction

### 1.1 The Global Fertility Paradox

Recent demographic data highlight a pronounced collapse in reproductive output across East Asia and the West. China's 2025 statistics reveal a 17% year-on-year decline in birth rates (National Bureau of Statistics of China 2026), with the total fertility rate estimated below 1.0 (Xu 2026). This is mirrored in South Korea, where the TFR reached a world-record low of 0.72 in 2023 (Statistics Korea 2024), and in Japan, which struggles to maintain a rate of 1.2 (Ministry of Health, Labour and Welfare 2024). Paradoxically, this collapse has persisted despite a decade of extensive pro-natalist interventions and ostensibly favorable structural shifts—including family-planning policy relaxation, government financial subsidies, declining housing prices, and the mainstream rise of gender-egalitarian norms. Simultaneously, fertility rates across all major economies—from the traditionally pro-natalist Middle East to emerging India—have synchronized below the 2.1 replacement line (United Nations 2024).

This synchronization reveals that conventional demographic models do not fully capture the complex socio-cultural dynamics of the 21st century. Specifically, this sudden contraction cannot be solely attributed to macroeconomic stagnation, as aggregate growth trajectories across East Asia saw no commensurate structural break over the past decade. The divergence of standard projections indicates that direct linear extrapolations of population change are structurally insufficient. Instead, fertility must be modeled through non-linear dynamics that capture the microscopic interactions of male-female marriage, thereby revealing the mechanisms driving phase transitions in birth rates in modern digital societies. By a phase transition, we mean a system that experiences no significant external shocks but demonstrates a sudden change due to the accumulation of small shifts—such as the recent collapse of the Chinese birth rate.

### 1.2 The Mechanics of Human Dating and Mating

From a mechanistic perspective, individual fertility outcomes depend primarily on the factors directly acting upon the agent. We argue that the major driving force for marriage is the initiation of a romantic relationship, which is a pair-bonding process. Based on psychology, the drive to initiate contact is often decoupled from long-term economic barriers, such as the lifetime cost of child-rearing, as individuals do not consider whether to have a child when they first identify someone they like. Instead, the primary determinant at this stage is the woman's expectation and acceptance during the initial encounter—whether through family arrangements, random meetings, or dating apps (Oppenheimer 1988; Raymo et al. 2015). Traditional economic models pass over this initial contact process and thus ignore the dynamics of what drives a relationship from dating to marriage, reducing the model to a linear correlation between birth rates and macroeconomic factors. When the intermediate transmission mechanism—the psychological drive toward romantic relationships—is disrupted by technological and social interventions and becomes the rate-limiting step, these traditional predictive structures fail.

## 1.3 Technological Disruption of the Libidinal Economy

In the last decade, widespread technology has fundamentally altered the mating mechanism. First, the smartphone's impact on the dopaminergic system has been linked to a reduced tolerance for delayed gratification and long-term investment (Lembke 2021). Second, the digital environment amplifies anxiety by exposing users to a curated "hyper-reality"—global lifestyles representing the top 0.1% of wealth and status (Baudrillard 1994). This induces a state of relative deprivation (Festinger 1954; Macunovich 1998), causing people's expectations for their future lives to rise disproportionately to local realities, while the perceived inadequacy that upward comparison produces (Vogel et al. 2014) depresses male mating interest. The same asymmetry is visible on dating platforms, now the most common venue for meeting a partner (Rosenfeld et al. 2019), where women accumulate matches rapidly while men accumulate them slowly (Tyson et al. 2016). This trend is poised to worsen as AI evolves into an "emotional assistant," offering synthetic intimacy that increasingly substitutes for interpersonal courtship. This phenomenon is not unique to humans. For example, humans have long applied mating disruption techniques in insect control by releasing synthetic mating signals (Cardé and Minks 1995). Similarly, AI and modern technology are now acting as "synthetic mating disruptors," analogous to pheromonal disruption in ecological population control.

## 1.4 Existing Non-linear Modeling Methods in Social Science

To account for these complex factors, researchers have consistently adapted non-linear modeling methods from physics and other quantitative fields into the social sciences. The application of non-linear modeling to social science is already well-established (Castellano et al. 2009; Galam 2012), providing a methodological foundation for using physical frameworks to understand complex human behaviors. Previous research has used statistical mechanics and many-particle assumptions to model income distribution—most notably in the work of Dragulescu and Yakovenko (2000), reviewed by Yakovenko and Rosser (2009) and Chakrabarti et al. (2013)—as well as wealth inequality and housing bubbles. These studies indicate that treating social agents as interacting particles subject to thermodynamic constraints is both feasible and empirically parsimonious across socioeconomic distributions. One strand is directly relevant here: Dil and Dil (2023) show that a deformed Fermi-Dirac distribution fits national income data more closely than Boltzmann-Gibbs forms in high-inequality systems. That work applies these statistics to the distribution of income itself; we instead use that distribution as the input spectrum for a selection process.

Given the strong correlation between economic status and marriage prospects, these sociophysical methods can be naturally extended from pure wealth distribution to the dynamics of the marriage market. This extension is crucial because existing demographic models often fail to capture the non-linear "phase transitions" that occur when a social system reaches a critical saturation of technological disruption (Scheffer et al. 2009). By treating mate selection as a process constrained by these underlying economic distributions, this study aims to provide the framework necessary to understand sudden fertility collapses.

## 1.5 Theoretical Approach and Contribution

This study bridges the gap between demographic observation, the mating mechanism, and sociophysical modeling. While demographic models include both linear and non-linear approaches, capturing how technological disruptions alter initial courtship—and how millions of individual pair-bonding decisions aggregate into a macro-level collapse—requires treating the marriage market as a many-particle system subject to thermodynamic constraints. By modeling human mating dynamics through Fermi-Dirac statistics—using the national income distribution as the input spectrum and female expectation as the threshold barrier—we quantify the "phase transition" occurring in global fertility. This paper contributes to the literature by: (1) proposing a novel "Marriage Energy Barrier" model that captures non-linear threshold dynamics missed by linear economic correlations; (2) quantifying the "Tech Factor" that distorts this barrier; and (3) projecting the accelerating impact of AI on future birth rates. This framework offers an alternative account of why financial incentives have failed to reverse the demographic collapse.

## 2. Applying Fermi-Dirac Models for Birth Rate Prediction

To model the fertility transition, we employ a non-linear aggregate model that isolates the core correlation between average income and fertility rates. Although cultural norms and policy shifts introduce variability across regions, we contend that the underlying threshold dynamics of mating are broadly generalizable across modernizing economies (United Nations 2024). To empirically validate this approach without the interference of transient local factors, we select Japan as our primary case study. Japan's relative cultural homogeneity and prolonged economic stagnation provide a stable "sociophysical equilibrium," making it a suitable environment to isolate the fundamental non-linear mechanics of the mating market.

To distinguish real systemic patterns from mere curve fitting, we adopt the following validation protocol. We calibrate the model parameters on a single year's cross-section, anchor the level at 2012, and then project fertility rates for the following decade. By using the subsequent years as a verification set, we show that our minimal-parameter model closely tracks the long-term structural trajectory of the mating market.

### 2.1 Marriage as the Structural Gatekeeper

Traditional demography assumes that reducing the economic burden of child-rearing promotes fertility. We contend that this misidentifies the primary bottleneck. The direct determinant of fertility is mating motivation. In modern East Asian societies, economic status serves as an indirect filter for marriage, but it is not the direct filter for procreation within a marriage. For instance, despite the sudden decrease in marriage rates in Japan, the average number of children born to married women remains near 2.0 (National Institute of Population and Social Security Research 2023). This indicates that annual birth counts are linearly coupled with annual marriage counts ($B = k \cdot M + c$), making marriage the primary gatekeeper for population replacement (Figure 1).

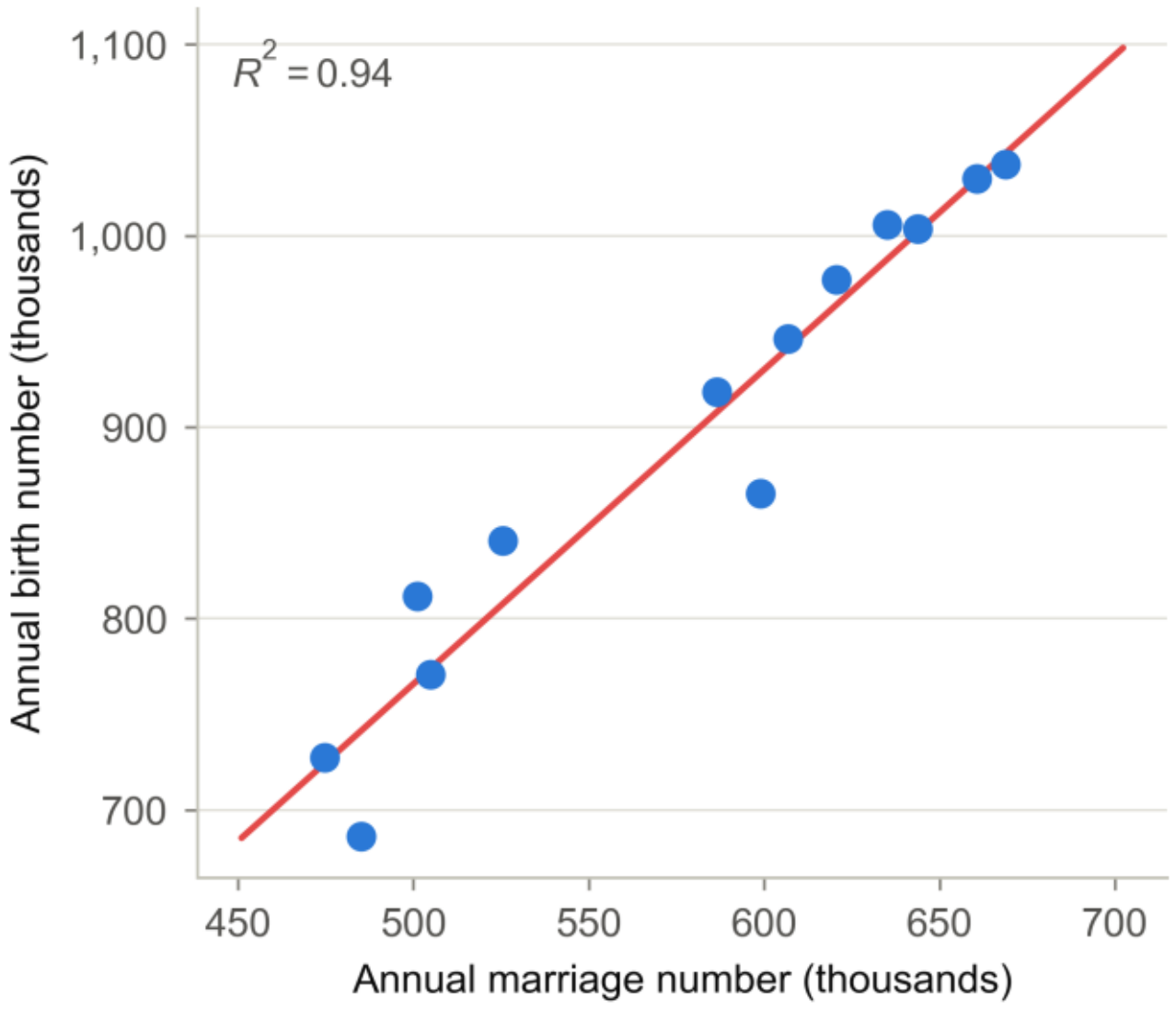

**Fig. 1** Linear correlation between annual marriage numbers and birth numbers in Japan ($R^2 = 0.94$)

### 2.2 Modeling the Marriage Energy Barrier

Having verified that annual births exhibit a strong linear dependency on marriage numbers, we can effectively reduce the problem of predicting fertility to that of predicting marriage entry. To quantify the impact of modern socio-technical factors, we propose that a large human population can be treated as a thermodynamic ensemble of interacting particles. Within this system, the pair-bonding between a man and a woman is modeled as a chemical interaction that must overcome a specific financial threshold, or "Energy Barrier". Chemistry calls this the activation energy: the push a reaction needs to get started, which can be far larger than whatever the reaction saves once it is done.

In East Asian cultures, women typically hold the primary discretion for marriage entry due to cultural norms and supply-side limitations. Thus, modeling marriage entry as a threshold evaluation in which prospective male partners are assessed against women's economic expectations is a reasonable structural assumption grounded in East Asian demographic realities (Raymo et al. 2015). This holds whether or not a woman is economically independent. Where she decides for herself, a man is judged against the standard she holds; where her family takes part in the decision, he is judged against requirements it states explicitly, as the Japanese and Chinese cases below illustrate. The gate is in the same place in both, and it asks the same question—whether a man can support a household—a judgement in which his income is a substantial part. The model therefore carries no separate terms for city and countryside, or for a woman's own earnings. It also follows that the standard can rise without anyone's preferences changing: the digital environment described in Section 1.3 reaches whoever applies the test as readily as the man being assessed. For a man, crossing this threshold requires meeting a substantial financial "activation energy" that is distinct from and often exceeds the daily cost of living. In Japan, the initial outlay associated with marriage entry (ceremony, gifts and setting up a household) is commonly put at 3 to 4 million JPY, approaching an entire year's salary for the average man in addition to providing better housing after marriage. In China, the barrier is even more rigid, often requiring real estate ownership and significant bride prices (Chae and Zhang 2026). These one-time outlays motivate the activation-energy analogy; the model evaluates the annual income a man must earn to make such an outlay affordable. The financial threshold required to *enter* marriage is now perceived as far higher than the ongoing cost of living either as a single individual or as a married couple. Indeed, in most modern economies, the per-capita living cost of a married couple is lower than that of a single individual due to economies of scale in shared housing and consumption (Figure 2).

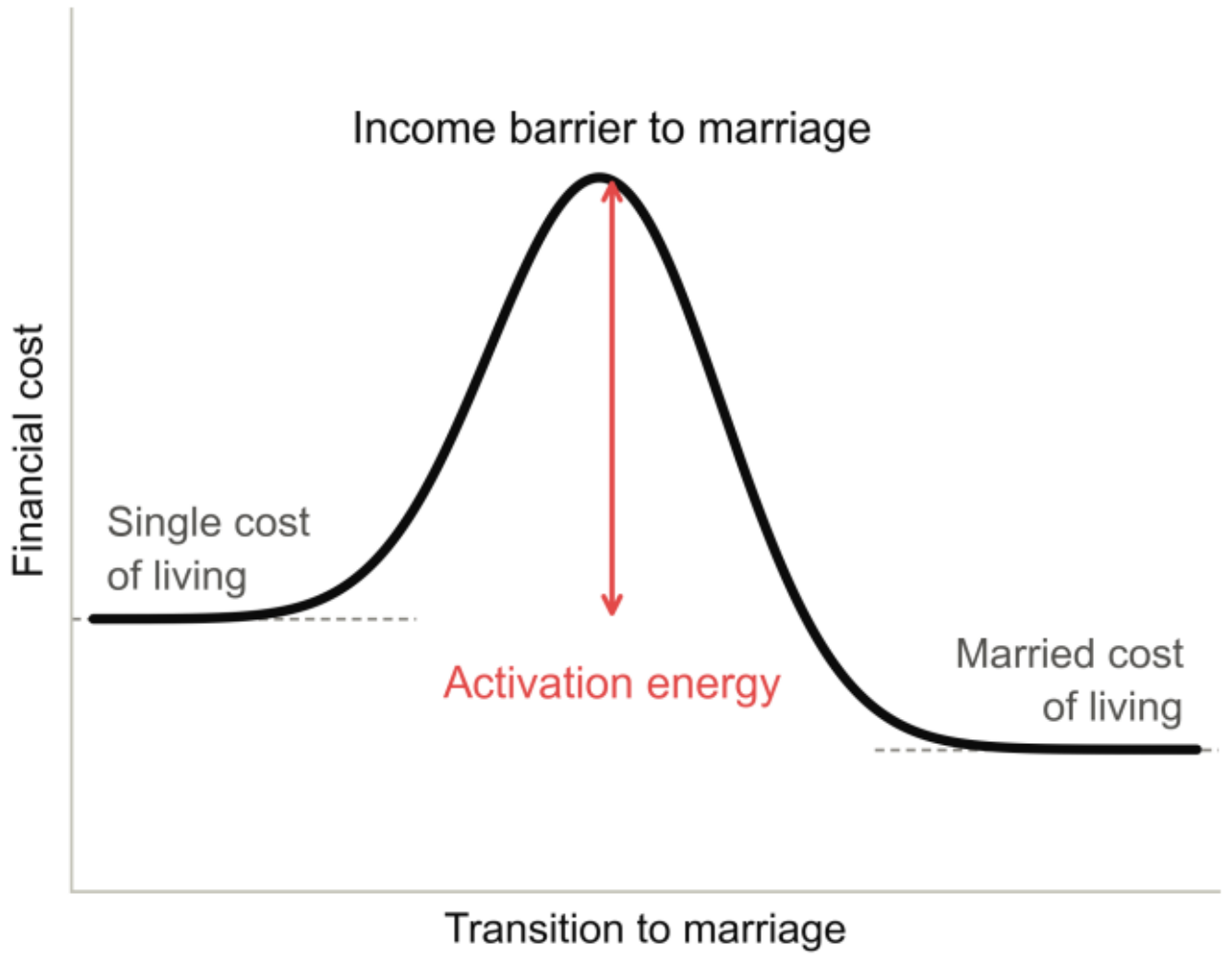

**Fig. 2** Proposed male income barrier acting as the sole determinant of marriage and fertility. The schematic illustrates the "Activation Energy" required to transition from Single Cost of Living to Married Cost of Living

Based on these assumptions, we propose two non-linear equations expressing the probability of marriage as a function of male income.

*The Arrhenius Application:* Expressed in terms of the ratio of required threshold income ($I_{barrier}$) to actual male income (I), the marriage probability P(I) across the male population can be initially approximated using an Arrhenius-type activation equation:

$$P(I) = \alpha \exp\left(-\frac{I_{\text{barrier}}}{I}\right) \tag{1}$$

where $I_{barrier}$ represents the mean annual male income that women expect of a prospective partner, and α is a scaling pre-factor. Equation (1) illustrates why marriage rates plummet exponentially when actual income I fails to keep pace with algorithmically inflated expectations ($I_{barrier}$): for low-income men, whose income sits far below $I_{barrier}$, marriage probability is negligible, whereas crossing the barrier triggers a sharp non-linear increase.

**The Fermi-Dirac Improvement:** However, the Arrhenius formulation approaches α as income grows, so the fitted α of 1.4578 lets it return P(I) > 100% at high incomes, which is physically unrealistic. To account for high-income saturation—where marriage probability asymptotes to a ceiling $P_{max}$ among top earners—we reformulate the transition kernel using Fermi-Dirac statistics (Dil and Dil 2023). The choice of Fermi-Dirac over Bose-Einstein or Maxwell-Boltzmann statistics is dictated by the structure of the market itself: under monogamy an agent occupies at most one marital state, an occupancy limit of unity that is the social analogue of the Pauli exclusion principle:

$$P(I) = \frac{P_{\max}}{1+\exp\left[\frac{I_{\text{barrier}}-I}{\sigma}\right]} \tag{2}$$

In this refined model, σ represents the "social temperature", the parameter setting how widely mating standards are spread across the population, and the ceiling $P_{\max}$ is fixed at 0.95. Equation (2) parsimoniously captures the macroscopic phase transition: as digital hyper-reality inflates $I_{\text{barrier}}$ while real male income I stagnates, the bulk of the male income distribution is shifted below the chemical potential ($I < I_{\text{barrier}}$) into the "forbidden" region of the Fermi-Dirac curve, precipitating a non-linear collapse in aggregate marriage and birth rates. In plainer terms, the chemical potential is the income at which a man's chance of marrying is half of $P_{\max}$. If everyone applies much the same standard, σ is small and that chance changes sharply as income crosses the threshold; if standards vary a great deal, σ is large, the change is gradual, and some men earning less than the barrier still marry.

## 2.3 Model Verification with Japanese Data

To verify the effectiveness of the models, we analyzed birth rates, marriage statistics, male population distribution, and birth cohort data (lagged by 32 years due to the average male marriage age) obtained from official Japanese Government statistics for the period 2012–2024. Dividing one year's marriages by the births of 32 years earlier gives the share of a single birth cohort that married that year (Figure S1; the lag itself is varied in Figure S2; all S-numbered figures are in Online Resource 1), which keeps that share comparable even as the population shrinks. The cutoff year of 2012 was selected as it marks the stabilization period following the 2008 financial crisis and the threshold where smartphone penetration exceeded 20%. The theoretical $I_{\text{barrier}}$ in this case is assumed to be 1.1 times the average income of all men, which aligns with the economic expectations unmarried Japanese women report bringing to marriage. In a national survey analysis, a partner's finances were rated important or worth considering by 94% of women against 41% of men, and about nine in ten women preferred a partner earning no less than they do (Ghaznavi et al. 2022). We further obtained the marriage rate of one anonymous Japanese city from a recent comprehensive study of administrative tax records reported by the University of Tokyo (Masaki 2024). If the model accurately reproduces this cross-sectional municipal distribution using only a baseline calibration of average income, it supports the empirical plausibility of our structural assumptions (Figure 3; see Figure 4a for the salary trend).

To determine the fitting factors in these refined models, we first estimated the effective total marriage rate compared to the number of cohort births 32 years prior, finding that the average marriage rate is approximately 40%. We then used the 2024 marriage rate data to determine the fitting parameters by constraining the total marriage rate to match actual data. This yielded a fitting parameter of α = 1.4578 for the Arrhenius model and a Fermi-Dirac fitting parameter of σ = 2.40 million JPY. Figure S3 sets out the identification of σ.

With these fitting parameters, the Fermi-Dirac model closely matched municipal data reported by the University of Tokyo (Masaki 2024). The data showed an estimated marriage rate lower than 10% for incomes under 2 million JPY, which rose to more than 55% when the income surpassed the average. Through this single-parameter calibration—conducted independently of the longitudinal test data—our Fermi-Dirac formulation captures the non-linear empirical structure of the Japanese mating market. This indicates that the observed cross-sectional marriage distribution in Japan is well approximated by the Fermi-Dirac functional form (Table 1).

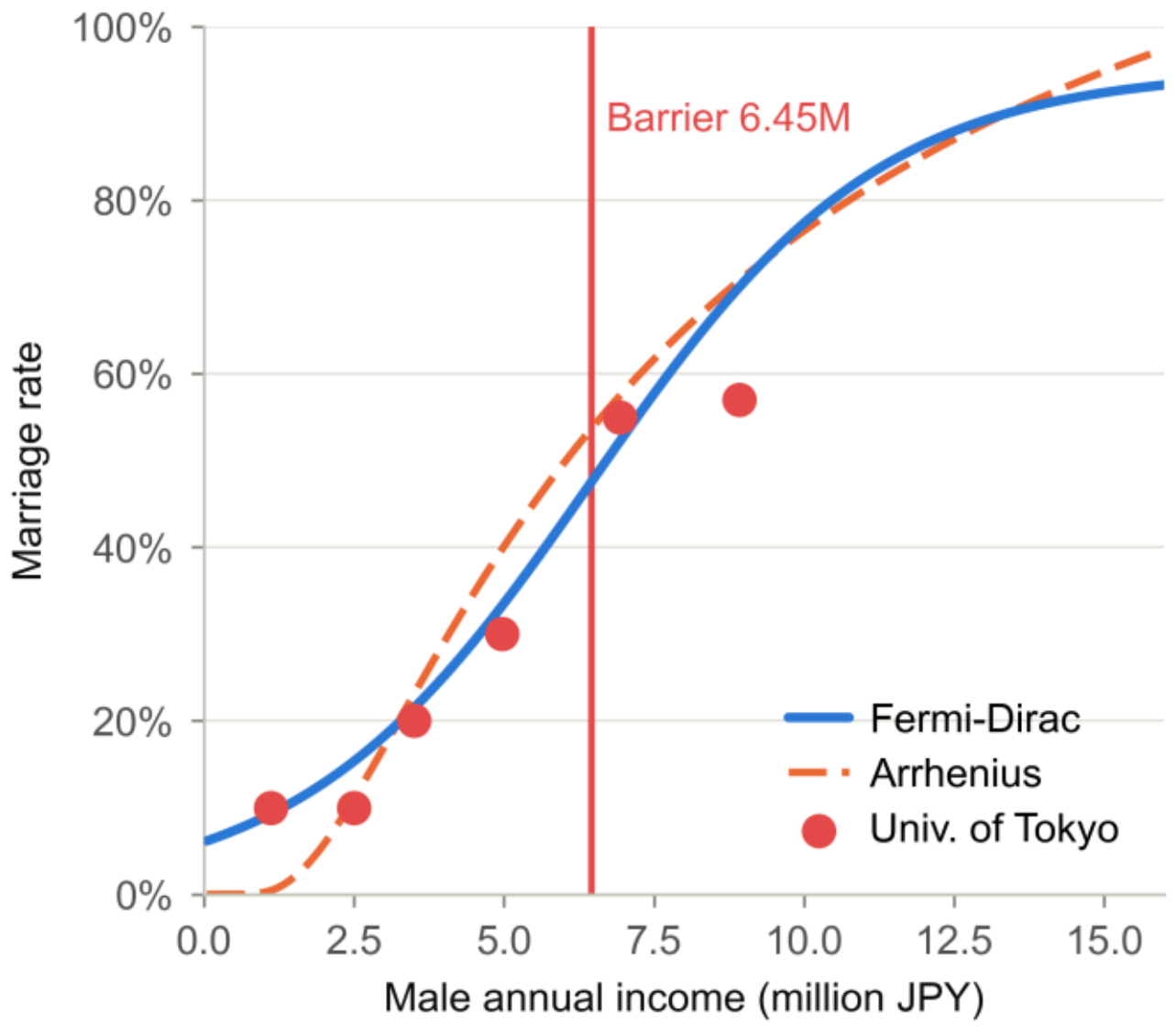

**Fig. 3** 2024 Marriage Probabilities by Income: Model Predictions vs. University of Tokyo Reference Data. The Fermi–Dirac model (blue line) demonstrates a superior fit to the reference data points (red dots) compared to the Arrhenius model (orange line), particularly in the transition zones. The barrier is 6.45 million JPY

**Table 1** Comparison of Model Predictions vs. Actual Marriage Rates in Japan (Year: 2024; Barrier: 6.45 M JPY; Actual Aggregated Rate: 40.12%). Difference vs Actual (Fermi): −0.04 percentage points.

| Band (JPY) | Share (%) | Mean income (M) | Fermi-Dirac | Arrhenius | Reference |
|---|---|---|---|---|---|
| Under 2.0M | 9.1 | 1.12 | 9.27% | 0.45% | 10% |
| 2.0M - 3.0M | 8.7 | 2.50 | 15.34% | 11.03% | 10% |
| 3.0M - 4.0M | 14.3 | 3.50 | 21.48% | 23.06% | 20% |
| 4.0M - 6.0M | 31.6 | 4.97 | 33.22% | 39.74% | 30% |
| 6.0M - 8.0M | 17.9 | 6.92 | 52.14% | 57.40% | 55% |
| 8.0M - 10.0M | 8.6 | 8.92 | 69.95% | 70.70% | 57% |
| Over 10.0M | 9.7 | 14.87 | 92.23% | 94.46% | – |
| **Total** | **99.9** | – | **40.08%** | **42.42%** | **40.12%** |

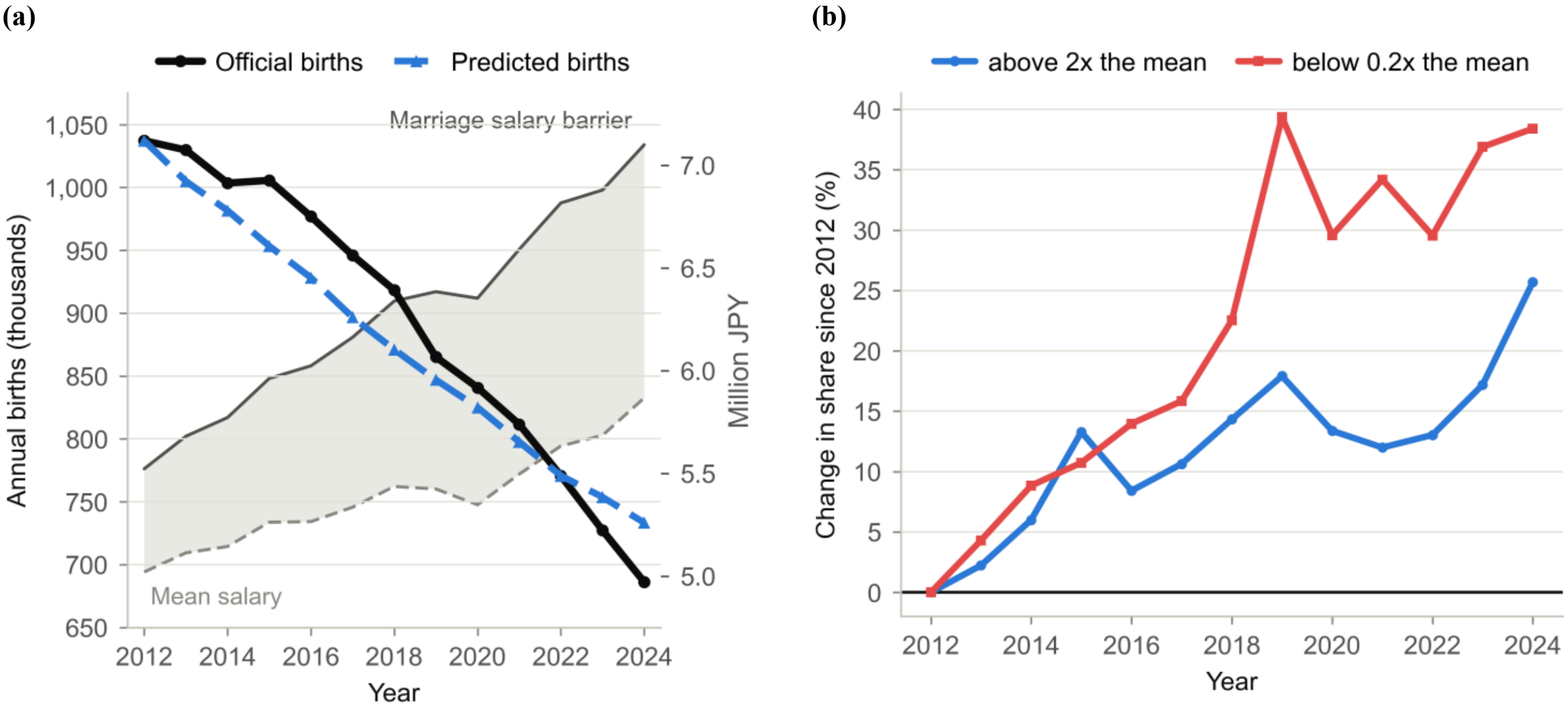

**Fig. 4** Birth predictions and the income data behind the barrier. **(a)** Historical validation of the Tech-only model, 2012–2024 (Table 2). Official and predicted annual births on the left axis; mean male salary and the marriage salary barrier on the right. Mean salary rises from 5.02 to 5.87 million JPY while the barrier rises from 5.52 to 7.10 million, so the shaded gap between them widens from 0.50 to 1.23 million. **(b)** Male earnings polarisation over the same window (National Tax Agency, composition of salaried earners, male). Each year's distribution is cut at multiples of that same year's mean male salary, so the bands do not drift with nominal wage growth. The share of men earning above twice the mean rises by 26% and the share earning below a fifth of the mean by 38%: the distribution widens at both ends at once. The widening may itself be driven by the Tech Factor, since it is the top of the distribution that is displayed, so the income that reads as adequate is set by what is shown off rather than by what is typical

## 2.4 Modeling Historical Birth Trends and the "Tech Factor" (2012–2024)

We next assessed the model's predictive accuracy against birth rate data from the past 12 years. We used official government data on male income distribution (Figure S4) to estimate marriage probabilities across different income brackets using our Fermi-Dirac model. These probabilities were weighted by the population distribution to derive an effective marriage rate for the entire male cohort. Total birth numbers were then calculated by multiplying the effective marriage rate by the population size and a linear fertility factor (derived from Figure 1), calibrated using 2012 baseline data.

In our predictive framework, we hypothesize that technology suppresses birth rates by inflating the "expected income" of an ideal partner—a result of social media exposing users to the hyper-curated lifestyles of the top 0.1%. To quantify this, we introduced a *Tech Factor* multiplier to the income barrier ($I_{barrier}$) term. This factor acts as a proxy for the increasing saturation of digital hyper-reality, mirroring the steady increase in daily smartphone usage hours observed in modern economies. Social media shows its users a misleading sample of how other people live, so the income expected of a partner climbs even when local wages do not. What has changed is the standard, not anyone's pay, so the model raises the barrier and leaves the measured income distribution alone.

The Tech Factor is initialized at 1.0 in 2012 (implying no distortion) and increases linearly by 0.01 annually, capping at 1.10. This represents a subtle, cumulative increase in expectations—approximately 1% of the average salary per year—which remains almost imperceptible to the individual (e.g., a 600,000 JPY increase on a 6 million JPY salary over a decade). Crucially, this idealized linear increment is designed to show how an almost imperceptible annual drift in expectations (1%/year) can trigger a macroscopic phase transition in fertility when the population sits near the Fermi-Dirac threshold. Although one could numerically optimize this annual growth rate to minimize longitudinal error over the 2013–2022 window (which yields an optimum near 0.65%/year), doing so would constitute data leakage from the test set into the calibration phase. To keep the longitudinal window free of post-hoc optimisation, we fix the Tech Factor *a priori* to a round, un-fitted increment of +0.01/year (capped at 1.10).

By incorporating this factor, the model reproduced birth numbers to within 5.21% in every year to 2022 (Table 2; Figure 4a). This supports the model's robustness, achieving high accuracy through single-point calibration rather than overfitting. Given the kernel parameters identified from the 2024 cross-section in Section 2.3, only one further number is tuned to the longitudinal birth series, and it is set so that 2012 matches the record exactly. No later year of that series was used in choosing it.

**Table 2** Historical Validation of the Model (2012–2024). Tech Factor only; AI Factor excluded.

| Year | Males 25-39 (10k) | Tech Factor | Barrier (M JPY) | Predicted births | Actual births | Error (%) |
|---|---|---|---|---|---|---|
| 2012 | 985 | 1.00 | 5.52 | 1,037,231 | 1,037,231 | +0.00 |
| 2013 | 972 | 1.01 | 5.68 | 1,004,957 | 1,029,816 | -2.41 |
| 2014 | 958 | 1.02 | 5.77 | 981,669 | 1,003,539 | -2.18 |
| 2015 | 945 | 1.03 | 5.96 | 953,416 | 1,005,677 | -5.20 |
| 2016 | 930 | 1.04 | 6.02 | 928,278 | 976,978 | -4.98 |
| 2017 | 915 | 1.05 | 6.16 | 896,795 | 946,065 | -5.21 |
| 2018 | 900 | 1.06 | 6.34 | 871,130 | 918,400 | -5.15 |
| 2019 | 885 | 1.07 | 6.38 | 847,020 | 865,239 | -2.11 |
| 2020 | 870 | 1.08 | 6.35 | 825,092 | 840,835 | -1.87 |
| 2021 | 855 | 1.09 | 6.59 | 797,657 | 811,622 | -1.72 |
| 2022 | 840 | 1.10 | 6.82 | 770,751 | 770,759 | -0.00 |
| 2023 | 825 | 1.10 | 6.88 | 753,686 | 727,288 | +3.63 |
| 2024 | 810 | 1.10 | 7.10 | 733,316 | 686,173 | +6.87 |

# 3. The AI Event Horizon: Accelerating the Collapse (2023–2040)

## 3.1 The "AI Factor" and Hyper-Reality

While the Tech-only model tracked observed Japanese births to within 5.21% from 2012 through 2022, a systematic divergence emerged in 2023 (+3.63%) and widened in 2024 (+6.87%; Figure 5). Although an error below 7% remains modest in absolute terms, the sudden directional break signals a secondary structural regime shift coinciding with the mass deployment of generative AI. We hypothesize that generative AI compounds the hyper-reality effect by simultaneously inflating lifestyle and partner expectations and providing synthetic emotional substitution (Malfacini 2025).

To quantify this regime shift, we introduce a multiplicative AI Factor into the effective barrier ($I_{eff} = I_{barrier} \times$ Tech × AI). Calibrating this term against the 2023 and 2024 residuals (Figure S5a–b) yields an annual growth rate of +0.025/year (capped at 1.25)—2.5 times the pace of the pre-AI Tech Factor. This steeper trajectory reflects the accelerated capacity of interactive AI systems to displace human courtship and emotional utility. Chatbots carry the same effect from pictures into conversation, so the inflated standard is now something a user talks to rather than only looks at. It is applied to the barrier in the same way as the Tech Factor.

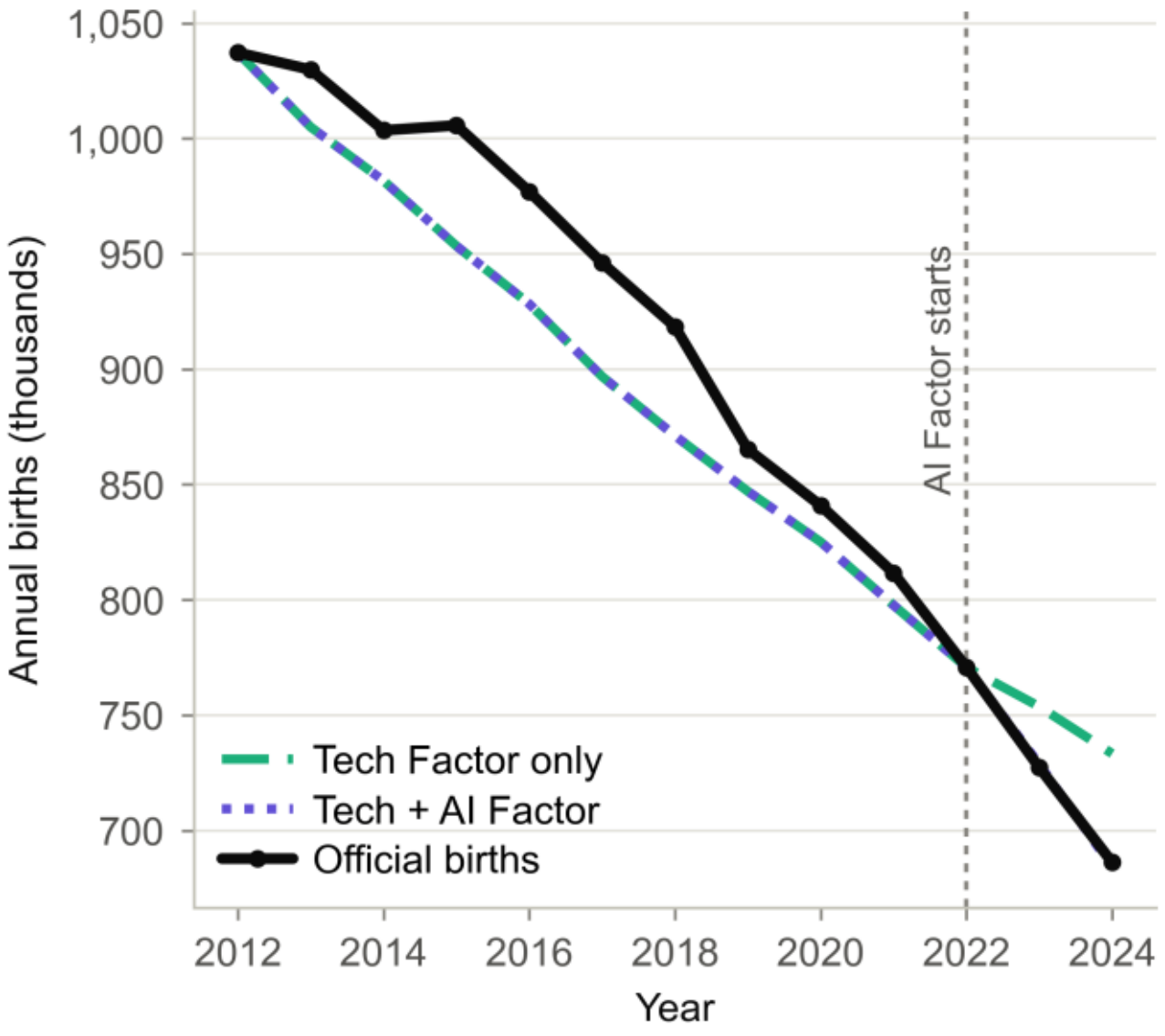

**Fig. 5** Japanese birth numbers against the model with the Tech Factor alone and with the Tech and AI Factors (2012–2024). The two runs are identical through 2022; the AI Factor first increases in 2023. The Tech Factor alone predicts 3.63% more births than occurred in 2023 and 6.87% more in 2024; with the AI Factor the same run gives 0.20% and -0.28%

### 3.2 Projections to 2040: The Extinction Spiral

Projecting this trend to 2040 reveals an outcome without precedent in Japan's modern record. As shown in Table 3, if the AI Factor continues to widen the gap between actual income and the "Effective Barrier" ($I_{barrier}$ × Tech × AI), annual birth numbers in Japan could plummet to approximately *208,000* by 2040 (Figure 6; 207,700 in Table 3). Section 4.4 discusses how this gap can widen when Mean Income rises while Median Income stagnates or grows more slowly. This projection assumes a linear growth cap, and Figure S6 varies the three assumptions it rests on; however, the accelerating pace of AI adoption suggests the impact could be even more severe.

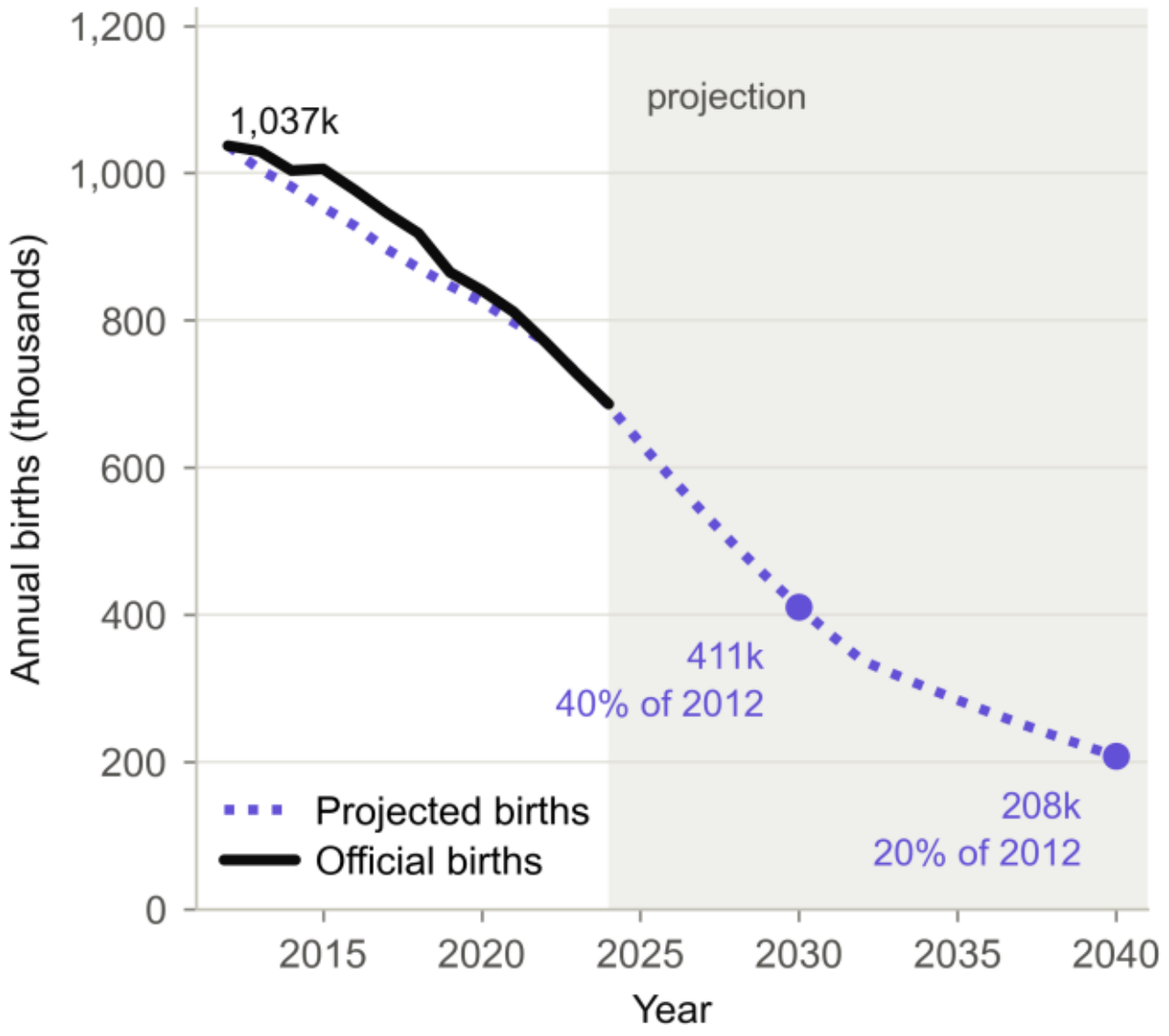

**Fig. 6** Forecast of Japanese birth numbers adjusting for AI acceleration (Table 3). Annual births fall to 410,702 by 2030 and to 207,700 by 2040, 40% and 20% of the 2012 level. The population scenario after 2040 is reported in Table S1

**Table 3** Forecast of Birth Numbers Adjusting for AI Acceleration (2023–2040)

| Year | Male Pop (10k, 25–39) | Tech Factor | AI Factor | Eff. Barrier (M JPY) | Predicted Births | Actual/Ref | Error % |
|---|---|---|---|---|---|---|---|
| 2023 | 825.00 | 1.10 | 1.025 | 7.05 | 728,735 | 727,288 | 0.20 |
| 2024 | 810.00 | 1.10 | 1.050 | 7.45 | 684,229 | 686,173 | -0.28 |
| 2025 | 795.00 | 1.10 | 1.075 | 7.75 | 633,402 | – | – |
| 2026 | 780.00 | 1.10 | 1.100 | 8.05 | 584,537 | – | – |
| 2027 | 765.00 | 1.10 | 1.125 | 8.35 | 537,750 | – | – |
| 2028 | 750.00 | 1.10 | 1.150 | 8.66 | 493,134 | – | – |
| 2029 | 735.00 | 1.10 | 1.175 | 8.99 | 450,767 | – | – |
| 2030 | 720.00 | 1.10 | 1.200 | 9.31 | 410,702 | – | – |
| 2031 | 705.00 | 1.10 | 1.225 | 9.65 | 372,971 | – | – |
| 2032 | 690.00 | 1.10 | 1.250 | 10.00 | 337,585 | – | – |
| 2033 | 675.00 | 1.10 | 1.250 | 10.15 | 319,104 | – | – |
| 2034 | 660.00 | 1.10 | 1.250 | 10.30 | 301,275 | – | – |
| 2035 | 645.00 | 1.10 | 1.250 | 10.45 | 284,093 | – | – |
| 2036 | 630.00 | 1.10 | 1.250 | 10.61 | 267,555 | – | – |
| 2037 | 615.00 | 1.10 | 1.250 | 10.77 | 251,654 | – | – |
| 2038 | 600.00 | 1.10 | 1.250 | 10.93 | 236,383 | – | – |
| 2039 | 585.00 | 1.10 | 1.250 | 11.09 | 221,734 | – | – |
| 2040 | 570.00 | 1.10 | 1.250 | 11.26 | 207,700 | – | – |

### 3.3 Underestimating the AI Multiplier: The Case of China

To test whether our projections are overly pessimistic, we performed an out-of-sample stress test by applying our model to China (2012–2025; National Bureau of Statistics of China 2025, 2026) using the exact same Tech and AI Factor schedules derived from Japan, without re-fitting the multipliers to Chinese fertility data (Figure 7; Tables 4 and 5; Figure S7). While a transient negative residual appears in 2016 (−15.44%) following the implementation of China's Universal Two-Child Policy, the model's predicted births in recent years are consistently *higher* than actual recorded births, with the over-prediction reaching +24.76% in 2025 (9.88M predicted vs. 7.92M observed; National Bureau of Statistics of China 2026).

Given China's hyper-digitalized mobile ecosystem and rigid marriage entry barriers (where urban real estate ownership and substantial bride prices elevate the baseline activation energy to 1.2× mean salary), this positive divergence suggests that our baseline multipliers are comparatively conservative. Rather than overstating the demographic threat, the baseline Fermi-Dirac parameterization *underestimates* the velocity of fertility collapse in societies experiencing rapid digital saturation and high baseline marriage barriers.

**Table 4** Model Validation Against Chinese Birth Data (2012–2025)

| Year | Pred_Births (M) | Actual_Births (M) | Error % | Avg_Salary (RMB) | Barrier (100k RMB) | Tech | AI |
|---|---|---|---|---|---|---|---|
| 2012 | 16.35 | 16.35 | 0.00 | 46,769 | 0.56 | 1.00 | 1.000 |
| 2013 | 16.01 | 16.40 | -2.37 | 51,474 | 0.62 | 1.01 | 1.000 |
| 2014 | 15.75 | 16.87 | -6.63 | 56,339 | 0.69 | 1.02 | 1.000 |
| 2015 | 15.42 | 16.55 | -6.80 | 62,029 | 0.77 | 1.03 | 1.000 |
| 2016 | 15.10 | 17.86 | -15.44 | 67,569 | 0.84 | 1.04 | 1.000 |
| 2017 | 14.61 | 17.23 | -15.18 | 74,318 | 0.94 | 1.05 | 1.000 |
| 2018 | 14.06 | 15.23 | -7.68 | 82,461 | 1.05 | 1.06 | 1.000 |
| 2019 | 13.55 | 14.65 | -7.53 | 90,501 | 1.16 | 1.07 | 1.000 |
| 2020 | 13.12 | 12.00 | 9.35 | 97,379 | 1.26 | 1.08 | 1.000 |
| 2021 | 12.61 | 10.62 | 18.72 | 106,837 | 1.40 | 1.09 | 1.000 |
| 2022 | 12.01 | 9.56 | 25.63 | 114,029 | 1.51 | 1.10 | 1.000 |
| 2023 | 11.21 | 9.02 | 24.24 | 120,698 | 1.63 | 1.10 | 1.025 |
| 2024 | 10.56 | 9.54 | 10.68 | 124,110 | 1.72 | 1.10 | 1.050 |
| 2025 | 9.88 | 7.92 | 24.76 | 127,600 | 1.81 | 1.10 | 1.075 |

**Table 5** Forecast of Chinese Birth Numbers Adjusting for AI Acceleration (2023–2040), using the same table format as Table 3

| Year | Male Pop (M, 25–39) | Tech Factor | AI Factor | Barrier (100k RMB) | Predicted (M) | Actual (M) | Error % |
|---|---|---|---|---|---|---|---|
| 2023 | 163.00 | 1.10 | 1.025 | 1.63 | 11.21 | 9.02 | 24.24 |
| 2024 | 161.00 | 1.10 | 1.050 | 1.72 | 10.56 | 9.54 | 10.68 |
| 2025 | 158.00 | 1.10 | 1.075 | 1.81 | 9.88 | 7.92 | 24.76 |
| 2026 | 155.00 | 1.10 | 1.100 | 1.91 | 9.23 | – | – |
| 2027 | 152.00 | 1.10 | 1.125 | 2.01 | 8.62 | – | – |
| 2028 | 149.00 | 1.10 | 1.150 | 2.12 | 8.05 | – | – |
| 2029 | 146.00 | 1.10 | 1.175 | 2.23 | 7.51 | – | – |
| 2030 | 143.00 | 1.10 | 1.200 | 2.34 | 7.00 | – | – |
| 2031 | 140.00 | 1.10 | 1.225 | 2.46 | 6.53 | – | – |
| 2032 | 137.00 | 1.10 | 1.250 | 2.59 | 6.08 | – | – |
| 2033 | 134.00 | 1.10 | 1.250 | 2.67 | 5.88 | – | – |
| 2034 | 131.00 | 1.10 | 1.250 | 2.75 | 5.69 | – | – |
| 2035 | 128.00 | 1.10 | 1.250 | 2.83 | 5.50 | – | – |
| 2036 | 125.00 | 1.10 | 1.250 | 2.91 | 5.31 | – | – |
| 2037 | 122.00 | 1.10 | 1.250 | 3.00 | 5.12 | – | – |
| 2038 | 119.00 | 1.10 | 1.250 | 3.09 | 4.94 | – | – |
| 2039 | 116.00 | 1.10 | 1.250 | 3.18 | 4.77 | – | – |
| 2040 | 113.00 | 1.10 | 1.250 | 3.28 | 4.59 | – | – |

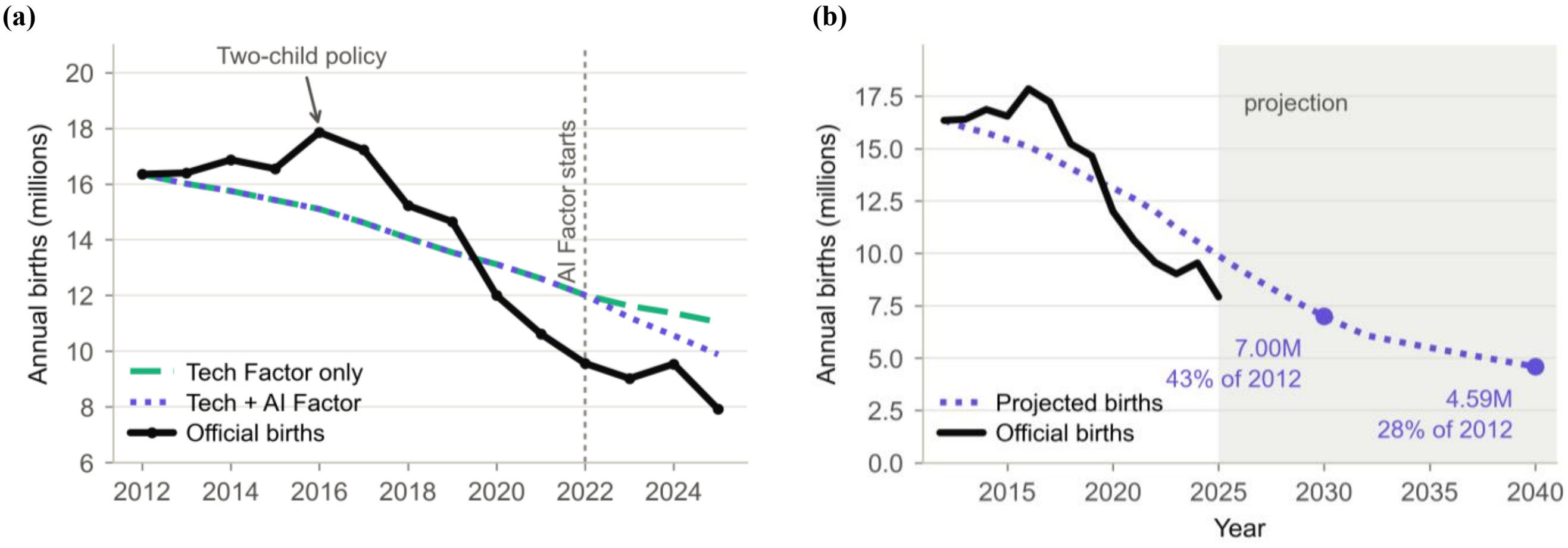

**Fig. 7** China Birth Rate Estimation: Barrier Model + Tech/AI Factors. **(a)** The model against official births, 2012–2025 (Table 4); the arrow marks the 2016 peak that followed the two-child policy. **(b)** The same model carried forward to 2040 (Table 5): 7.00 million births by 2030 and 4.59 million by 2040, 43% and 28% of the 2012 level. The population scenario after 2040 is reported in Table S2

# 4. Discussion: The Mechanics of Extinction

## 4.1 The AI Paradox: Emotional Displacement without Intelligence

A critique of our "AI Factor" might be that current AI (Generative Pre-trained Transformers) lacks true agency or intelligence. However, our analysis suggests that emotional appeal does not require human-like intelligence. As evidenced by user behaviors on platforms like Character.AI and Replika (Bakir and McStay 2025), current AI models are already sufficient to serve as "super-stimuli" for human intimacy, in the sense Tinbergen (1951) gave the term and Barrett (2010) extended to synthetic media. They offer validation without vulnerability and companionship without compromise (Skjuve et al. 2021; Pentina et al. 2023).

The danger is that this displacement is occurring *before* the arrival of Artificial General Intelligence (AGI). We are witnessing the onset of this spiral driven merely by advanced chatbots. If AGI were to become a reality—possessing the ability to not only simulate affection but to proactively anticipate and fulfill complex psychological needs—the rate of human displacement from the mating market could accelerate non-linearly. The "AI Factor" in our model could then shift from a linear growth (0.025/year) toward a compounding trajectory.

This population decline permanently reduces the cohort of women of reproductive age after a 20- to 30-year lag. Because human generational turnover is slow, population recovery after 2040 would require an unattainable surge in TFR (Figure 8b), exponentially increasing the difficulty of policy intervention and creating a point of no return driven by the non-linear phase transition.

## 4.2 The Global Synchronization of Decline

Traditionally, Demographic Transition Theory posits that fertility declines lag significantly behind economic development, occurring only after societies reach a certain threshold of industrialization, education, and health. However, the current global demographic landscape is characterized by a "digital synchronization" that appears to bypass these economic stages entirely. We observe non-linear fertility collapses in developing and traditionally pro-natalist nations that have not yet reached the wealth levels of West or East Asia, suggesting that the information environment may be a more potent variable than GDP per capita (Vollset et al. 2020).

In emerging economies such as *Vietnam* and *Mexico*, the fertility trajectory has decoupled from the classic industrialization timeline. Vietnam, despite its lower per capita income relative to Western nations, has seen its Total Fertility Rate (TFR) slide below the replacement level to approximately 1.9, with urban centers like Ho Chi Minh City mirroring the ultra-low rates of East Asia (United Nations 2024). Similarly, Mexico has experienced a precipitous contraction, with recent estimates placing its TFR near 1.6, a figure that took European nations decades to reach after achieving significantly higher standards of living (United Nations 2024).

This phenomenon is perhaps most consequential in *India*, widely anticipated to be the global demographic engine of the 21st century. Contrary to projections of a prolonged "demographic dividend," India's national TFR has already fallen below the replacement threshold of 2.1 (International Institute for Population Sciences and ICF 2021). The ubiquity of affordable mobile data (the "Jio effect") has accelerated the diffusion of global lifestyle standards, coinciding with a rapid stratification of the mating market in urban India, effectively bypassing the gradual secularization phase observed in the West.

Furthermore, the influence of the digital information environment appears to transcend cultural and religious barriers. *Iran*, despite a religious governance framework that actively promotes traditional family norms, has recorded one of the fastest fertility collapses in recorded history. Having fallen from over 6.0 in the 1980s under state family planning, Iran's TFR has continued down to approximately 1.6 (United Nations 2024) despite a comprehensive reversal toward pro-natalist policy in the 2010s, indicating that state ideology and religious conservatism are insufficient insulators against the hyper-reality of the global digital sphere.

## 4.3 The Irreversible Loss of Relational Competence

Perhaps the most alarming implication of our findings is the potential irreversibility of this trend. Mating and caregiving are not merely biological instincts but learned social algorithms reinforced by culture. As a generation matures relying on AI for intimacy ("frictionless relationships"), they risk permanently losing the resilience required for human friction—the compromises and conflicts inherent in mating and child-rearing (Turkle 2011; Zimmerman et al. 2023).

This loss of competence coincides with a burgeoning mental health crisis. The reliance on AI companions has been linked to increased isolation and, paradoxically, to youth suicide in case reports involving anthropomorphic chatbots (Bakir and McStay 2025). Furthermore, the demographic collapse creates a secondary mortality crisis: the "caregiving vacuum." As the population of young people shrinks, the elderly will face a future of automated but insufficient care. Contrary to the techno-optimist view that technology extends life, severe strain on the human caregiving structure could place downward pressure on old-age health outcomes, creating a feedback loop that accelerates the depopulation process. We are not merely facing a decline in births, but a fundamental degradation of the human capacity to sustain life.

### 4.4 The Inequality Trap: Why AI-Driven Growth Will Not Rescue Fertility

A common counter-argument to demographic pessimism is the economic assumption that increased productivity driven by Artificial Intelligence will raise living standards and, consequently, fertility. Our model challenges this assumption by highlighting the divergence between *Mean* and *Median* income in an AI-dominated economy.

In our Fermi-Dirac framework, the "Activation Energy" required for marriage ($I_{barrier}$) is psychologically pegged to the *Mean Income* of the male population. This "Barrier" represents the social standard of a "viable provider," which is inflated by the "hyper-reality" of digital media. In an era of AI-driven wealth concentration, productivity gains accrue disproportionately to the top percentile of capital owners, driving the Mean Income upward significantly while the Median Income (representing the typical male earner) stagnates or grows at a slower rate due to labor displacement. This is the thermal bulk and super-thermal tail of the income spectrum pulling apart (Yakovenko and Rosser 2009; see Figure 4b for the Japanese earnings distribution).

Mathematically, this creates a "moving goalpost" dynamic. Defining the marriage probability $P_{median}$ for the median male earner ($I_{median}$) as:

$$P_{\text{median}} = \frac{P_{\text{max}}}{1+\exp\left[\frac{I_{\text{barrier}}-I_{\text{median}}}{\sigma}\right]} \tag{3}$$

As the spread between the Mean (which drives the Barrier) and the Median (the actual income) widens, the exponent becomes increasingly positive, driving the probability $P_{median}$ toward zero. Therefore, even if the aggregate economy grows ("rich people getting richer"), the *Effective Barrier* rises faster than the capacity of the majority to meet it. Consequently, AI-induced top-heavy economic growth acts as a structural suppressant of pair-bonding across the median population: the wealthier the society becomes at the top, the further below the expected threshold—and therefore the more structurally excluded from marriage entry—the median male earner becomes in the algorithmic mating market.

### 4.5 Western Stagnation as the "Lower Bound" of Human Reproduction

Our analysis purposefully prioritizes East Asian data over Western demographics. Critics may point to the exclusion of European data as a limitation; however, we contend that Western nations represent a "demographically dormant" state that records the floor this transition is moving towards rather than the active phase we set out to model.

Western Europe and North America have effectively operated below the replacement level for decades, maintaining population stability only through significant immigration flows. In thermodynamic terms, these regions represent a "cold reservoir" in which births alone no longer sustain the population. Set migration aside and the decline underneath it is steep enough that sustained sub-replacement fertility is forecast to carry economic, social and geopolitical consequences (Vollset et al. 2020). The difficulty for our purposes is one of measurement rather than of substance: headline population counts combine births with migration, so they cannot isolate the birth signal against which the "Tech Factor" is fitted.

Furthermore, these nations represent the "lower bound" of the extinction curve. Even with arguably higher levels of gender equality and social safety nets than East Asia, Western fertility remains stubbornly low. The Nordic countries are the clearest case. Long regarded as forerunners for their high and stable fertility alongside generous parental leave, subsidised childcare and high gender equality, they have seen a pronounced decline since around 2010 that is not merely postponement and that falls most heavily on first births, the entry step this paper models (Hellstrand et al. 2021). This suggests that cultural and policy interventions are insufficient to counteract the digital mating environment. If East Asia—currently the "hot reservoir" of the global industrial engine—undergoes the predicted phase transition and converges with the Western "lower bound," the global human population effectively loses its reproductive engine. The horizon for human population is not determined by the regions that are already relying on external inputs to survive, but by the rapid collapse of the regions that were once the primary drivers of human population growth.

## 5. Conclusion

### 5.1 The Digital Behavioral Sink

John B. Calhoun's "Universe 25" experiment (Calhoun 1973) has long been criticized by sociologists for its biological determinism (Ramsden and Adams 2009), on the grounds that human complexity and cultural adaptability preclude such a "behavioral sink" in our species. However, those experiments demonstrated that birth rates do not return to normal even when resources remain abundant and population density falls, suggesting that depopulation can become a self-reinforcing feedback loop. Thus, we cannot assume that higher productivity and a smaller population will automatically resolve the fertility crisis.

In addition, this paper suggests that Artificial Intelligence constitutes the final, missing variable in Calhoun's equation. By providing an unlimited supply of frictionless dopamine and synthetic validation, AI weakens the behavioral imperative for social cooperation, effectively completing the construction of a digital Universe 25.

The resulting dynamic is a dual-failure mode: a collapse of male mating motivation due to synthetic substitution, and a surge in reproductive anxiety among women driven by algorithmic hyper-reality. This leads to a state where the population ceases to grow not due to resource scarcity, but due to psychological withdrawal—a circumstance no less consequential than environmental climate change, and considerably less visible. While climate change degrades the habitat, the "Algorithmic Climate Crisis" erodes the biological will of the inhabitant to persist. The parallel extends to the lag. Warming already committed cannot be undone by later abatement, and a cohort not born cannot be restored once the interval has passed: on the 32-year cohort spacing used throughout this paper, the births forgone in 2025 are the prospective parents absent in 2057. Even a sudden recovery to a TFR of 1.0 after 2040 does not halt population decline in Japan or China (Figure 8; Tables S1 and S2).

**(a)** **(b)**

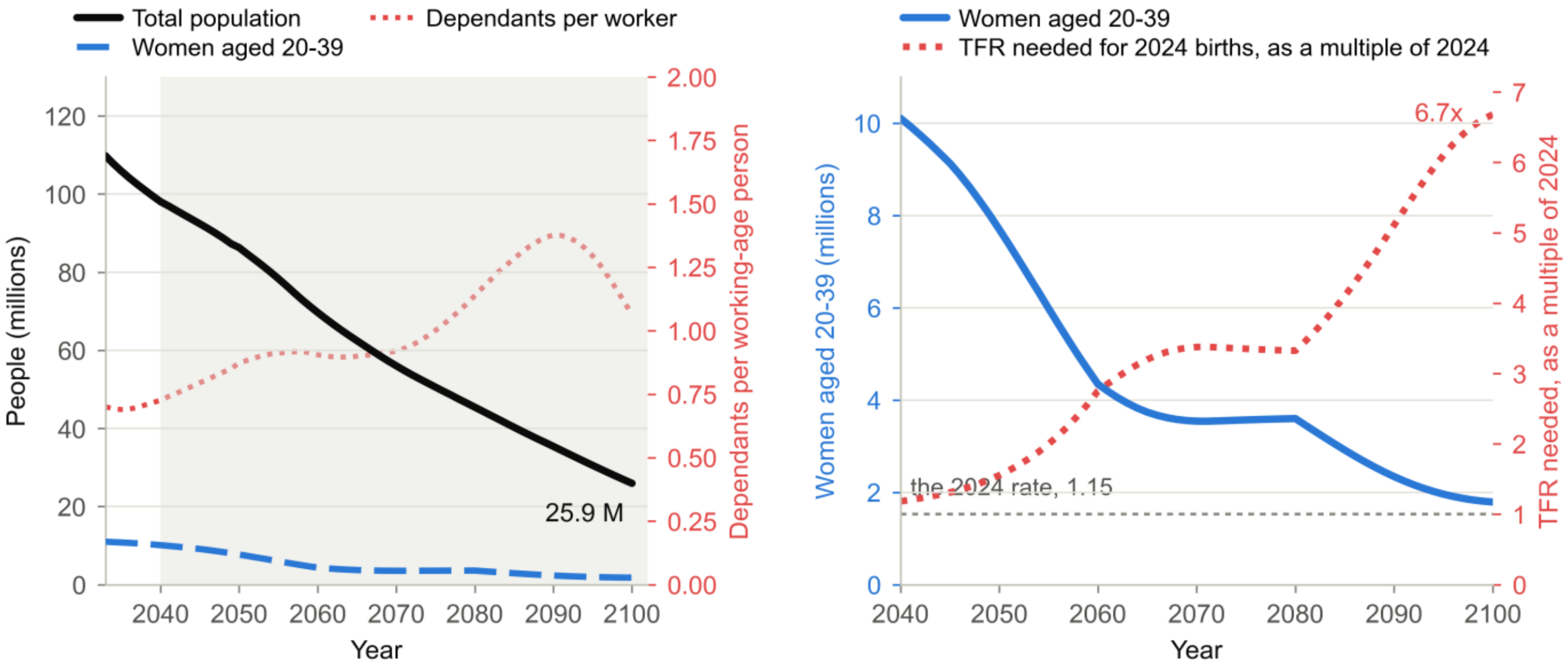

**Fig. 8** Japan after 2040 with the total fertility rate held at 1.0, its present order of magnitude. Births to 2040 are those of Table 3; the cohorts behind every quantity shown are already born. **(a)** Total population and women aged 20–39 on the left axis, dependants per working-age person on the right. The population falls from 98.1 million in 2040 to 25.9 million in 2100, and dependants exceed workers from the mid-2070s. **(b)** Women aged 20–39 against the total fertility rate that would be needed in each year to produce the 2024 birth number, expressed as a multiple of the 2024 rate of 1.15. The requirement rises from 1.18 times the 2024 rate in 2040 to 6.67 times by 2100, because the women who would bear those children are already born

## 5.2 The Great Filter and the Case for Prohibition

Viewed through the lens of the Fermi Paradox, AI-driven reproductive collapse may represent a "Great Filter"—a developmental trap where advanced civilizations engineer their own obsolescence through synthetic substitution before they can become multi-planetary. Consequently, we posit that standard regulatory frameworks are insufficient to address this threat.

If our projections hold, humanity faces a quantifiable trajectory toward extinction within the century: on the trajectory of Table 3, Japanese annual births fall to roughly 208,000 by 2040, little more than a quarter of the 2023 figure. The AI Factor stops rising when it reaches its ceiling in 2032, but the barrier it inflates goes on climbing with nominal pay. Every year of delay is a year in which the intervention required grows larger. To avoid this outcome, we argue for a drastic de-escalation of the attention economy. Absent stringent regulatory guardrails restricting the mass-market deployment of anthropomorphic AI designed for emotional and romantic substitution, our projections indicate that the downward phase transition in human fertility may become self-reinforcing and structurally irreversible by 2040. The factors that inflate the barrier are products of deployment decisions rather than of demography, and are open to policy in a way that an ageing population is not. We stand at a precipice where the convenience of the present is being purchased with the existence of the future.

## Declarations

**Use of Large Language Models** A large language model was used to revise and refactor the authors' initial analysis code. It changed no input, parameter, result or conclusion reported here, and the authors are accountable for the final version.

**Funding** The authors declare that no funds, grants, or other support were received during the preparation of this manuscript.

**Conflicts of Interest** The authors have no relevant financial or non-financial interests to disclose.

**Ethics Approval** This study utilized publicly available datasets from the Japanese and Chinese governments. No human participants or animals were involved in this study.

**Availability of Data and Materials** The datasets generated and/or analyzed during the current study are available from the Government of Japan Statistics Bureau and the National Bureau of Statistics of China.

**Code Availability** Code used for the Fermi-Dirac modeling, including the processed data used for the model fitting, is available from the corresponding author on request.

**Disclaimer:** This work is independent research and does not represent the views of Lawrence Berkeley National Laboratory (LBNL) or the U.S. Department of Energy (DOE).

## References

Bakir V, McStay A (2025) Move fast and break people? Ethics, companion apps, and the case of Character.ai. *AI & SOCIETY* 40(8):6365–6377. https://doi.org/10.1007/s00146-025-02408-5

Barrett D (2010) *Supernormal Stimuli: How Primal Urges Overran Their Evolutionary Purpose*. W. W. Norton, New York

Baudrillard J (1994) *Simulacra and Simulation* (trans: Glaser SF). University of Michigan Press, Ann Arbor

Calhoun JB (1973) Death squared: The explosive growth and demise of a mouse population. *Proceedings of the Royal Society of Medicine* 66(1 Pt 2):80–88. https://doi.org/10.1177/00359157730661p202

Cardé RT, Minks AK (1995) Control of moth pests by mating disruption: Successes and constraints. *Annual Review of Entomology* 40(1):559–585. https://doi.org/10.1146/annurev.en.40.010195.003015

Castellano C, Fortunato S, Loreto V (2009) Statistical physics of social dynamics. *Reviews of Modern Physics* 81(2):591–646. https://doi.org/10.1103/RevModPhys.81.591

Chae M, Zhang D (2026) Bride price, marriage expectations, and intentions among young male migrants in China: Evidence from manufacturing gig workers. *Economic Modelling* 155:107443. https://doi.org/10.1016/j.econmod.2025.107443

Chakrabarti BK, Chakraborti A, Chakravarty SR, Chatterjee A (2013) *Econophysics of Income and Wealth Distributions*. Cambridge University Press, Cambridge. https://doi.org/10.1017/CBO9781139004169

Dil E, Dil E (2023) Sociophysics of income distributions modeled by deformed Fermi-Dirac distributions. *The Journal of Mathematical Sociology* 47(2):97–122. https://doi.org/10.1080/0022250X.2021.1973456

Dragulescu A, Yakovenko VM (2000) Statistical mechanics of money. *The European Physical Journal B* 17(4):723–729. https://doi.org/10.1007/s100510070114

Festinger L (1954) A theory of social comparison processes. *Human Relations* 7(2):117–140. https://doi.org/10.1177/001872675400700202

Galam S (2012) *Sociophysics: A Physicist's Modeling of Psycho-political Phenomena*. Springer, New York. https://doi.org/10.1007/978-1-4614-2032-3

Ghaznavi C, Sakamoto H, Nomura S, Kubota A, Yoneoka D, Shibuya K, Ueda P (2022) Fish in the sea: Number, characteristics, and partner preferences of unmarried Japanese adults—Analysis of a national survey. *PLOS ONE* 17(2):e0262528. https://doi.org/10.1371/journal.pone.0262528

Hellstrand J, Nisén J, Miranda V, Fallesen P, Dommermuth L, Myrskylä M (2021) Not just later, but fewer: Novel trends in cohort fertility in the Nordic countries. *Demography* 58(4):1373–1399. https://doi.org/10.1215/00703370-9373618

International Institute for Population Sciences (IIPS), ICF (2021) *National Family Health Survey (NFHS-5), 2019–21: India*. IIPS, Mumbai. https://dhsprogram.com/pubs/pdf/FR375/FR375.pdf. Accessed 1 November 2026

Lembke A (2021) *Dopamine Nation: Finding Balance in the Age of Indulgence*. Dutton, New York

Macunovich DJ (1998) Fertility and the Easterlin hypothesis: An assessment of the literature. *Journal of Population Economics* 11(1):53–111. https://doi.org/10.1007/s001480050058

Malfacini K (2025) The impacts of companion AI on human relationships: Risks, benefits, and design considerations. *AI & SOCIETY* 40(7):5527–5540. https://doi.org/10.1007/s00146-025-02318-6

Masaki Y (2024) *Income and Fertility: Analyzing the Relationship Using Administrative Tax Records from a Japanese City*. GraSPP Discussion Paper E-24-001. Graduate School of Public Policy, The University of Tokyo. https://www.pp.u-tokyo.ac.jp/wp-content/uploads/2024/12/GraSPP-DP-E-24-001.pdf. Accessed 1 November 2026

Ministry of Health, Labour and Welfare (MHLW) (2024) *Vital Statistics of Japan*. Government of Japan, Tokyo. https://www.e-stat.go.jp/en/statistics/00450011. Accessed 1 November 2026

National Bureau of Statistics of China (2025) *China Statistical Yearbook 2025*. China Statistics Press, Beijing. https://www.stats.gov.cn/sj/ndsj/2025/indexeh.htm. Accessed 1 November 2026

National Bureau of Statistics of China (2026) *Statistical Communiqué of the People's Republic of China on the 2025 National Economic and Social Development*. National Bureau of Statistics of China, Beijing. https://www.stats.gov.cn/english/PressRelease/202602/t20260228_1962661.html. Accessed 1 November 2026

National Institute of Population and Social Security Research (IPSS) (2023) *The 16th Japanese National Fertility Survey*. IPSS, Tokyo. https://www.ipss.go.jp/ps-doukou/e/doukou16/Nfs16_gaiyoEng.html. Accessed 1 November 2026

Oppenheimer VK (1988) A theory of marriage timing. *American Journal of Sociology* 94(3):563–591. https://doi.org/10.1086/229030

Pentina I, Hancock T, Xie T (2023) Exploring relationship development with social chatbots: A mixed-method study of Replika. *Computers in Human Behavior* 140:107600. https://doi.org/10.1016/j.chb.2022.107600

Ramsden E, Adams J (2009) Escaping the laboratory: The rodent experiments of John B. Calhoun and their cultural influence. *Journal of Social History* 42(3):761–792. https://doi.org/10.1353/jsh.0.0156

Raymo JM, Park H, Xie Y, Yeung WJJ (2015) Marriage and family in East Asia: Continuity and change. *Annual Review of Sociology* 41:471–492. https://doi.org/10.1146/annurev-soc-073014-112428

Rosenfeld MJ, Thomas RJ, Hausen S (2019) Disintermediating your friends: How online dating in the United States displaces other ways of meeting. *Proceedings of the National Academy of Sciences* 116(36):17753–17758. https://doi.org/10.1073/pnas.1908630116

Scheffer M, Bascompte J, Brock WA, Brovkin V, Carpenter SR, Dakos V, Held H, van Nes EH, Rietkerk M, Sugihara G (2009) Early-warning signals for critical transitions. *Nature* 461(7260):53–59. https://doi.org/10.1038/nature08227

Skjuve M, Følstad A, Fostervold KI, Brandtzaeg PB (2021) My chatbot companion—a study of human-chatbot relationships. *International Journal of Human-Computer Studies* 149:102601. https://doi.org/10.1016/j.ijhcs.2021.102601

Statistics Korea (KOSTAT) (2024) *Preliminary Results of Birth and Death Statistics in 2023*. Statistics Korea, Daejeon. https://mods.go.kr/board.es?act=view&bid=11773&list_no=430941&mid=a20108100000. Accessed 1 November 2026

Tinbergen N (1951) *The Study of Instinct*. Clarendon Press, Oxford

Turkle S (2011) *Alone Together: Why We Expect More from Technology and Less from Each Other*. Basic Books, New York

Tyson G, Perta VC, Haddadi H, Seto MC (2016) A first look at user activity on Tinder. In: *Proceedings of the 2016 IEEE/ACM International Conference on Advances in Social Networks Analysis and Mining*. IEEE, pp 461–466. https://doi.org/10.1109/ASONAM.2016.7752275

United Nations, Department of Economic and Social Affairs, Population Division (2024) *World Population Prospects 2024*. United Nations, New York. https://population.un.org/wpp/. Accessed 1 November 2026

Vogel EA, Rose JP, Roberts LR, Eckles K (2014) Social comparison, social media, and self-esteem. *Psychology of Popular Media Culture* 3(4):206–222. https://doi.org/10.1037/ppm0000047

Vollset SE, Goren E, Yuan CW, Cao J, Smith AE, Hsiao T et al (2020) Fertility, mortality, migration, and population scenarios for 195 countries and territories from 2017 to 2100: A forecasting analysis for the Global Burden of Disease Study. *The Lancet* 396(10258):1285–1306. https://doi.org/10.1016/S0140-6736(20)30677-2

Xu W (2026) China's population decline accelerates as births hit modern-era low in 2025. *Caixin Global*. https://www.caixinglobal.com/2026-01-19/chinas-population-decline-accelerates-as-births-hit-modern-era-low-in-2025-102405521.html. Accessed 1 November 2026

Yakovenko VM, Rosser JB (2009) Colloquium: Statistical mechanics of money, wealth, and income. *Reviews of Modern Physics* 81(4):1703–1725. https://doi.org/10.1103/RevModPhys.81.1703

Zimmerman A, Janhonen J, Beer E (2023) Human/AI relationships: Challenges, downsides, and impacts on human/human relationships. *AI and Ethics* 4(4):1555–1567. https://doi.org/10.1007/s43681-023-00348-8

## Online Resource 1

Supplementary Information for “The algorithmic behavioral sink: Modeling the transition to irreversible demographic collapse”

Figures S1–S7 follow the order of the comparisons in the main text: cohort marriage rates and the cohort lag, social temperature, the male income distribution, the AI Factor, forecast scenarios, and the income reconstruction for China. Tables S1 and S2 give the annual population scenarios for Japan and China from 2040 to 2100. RMSE denotes root-mean-square error and pp denotes percentage points.

Note on calculations. The MHLW income-share tables sum to 99.9 rather than 100.0 in some years because of rounding; the cross-sectional check divides by the realised total and the birth-count run does not, following the original code in each case. Section 2.3, Table 1 and Figure 3 use the baseline barrier of 1.1 times mean male salary. The Tech Factor is introduced in Section 2.4 and the AI Factor in Section 3.1. Table 2 and Figure 4a use the Tech Factor alone; Table 3 and Figure 6 use both factors, while Figure 5 compares the two runs. The MHLW distribution used in the calculations is shown in Figure S4; the separate NTA distribution is shown in Figure 4b.

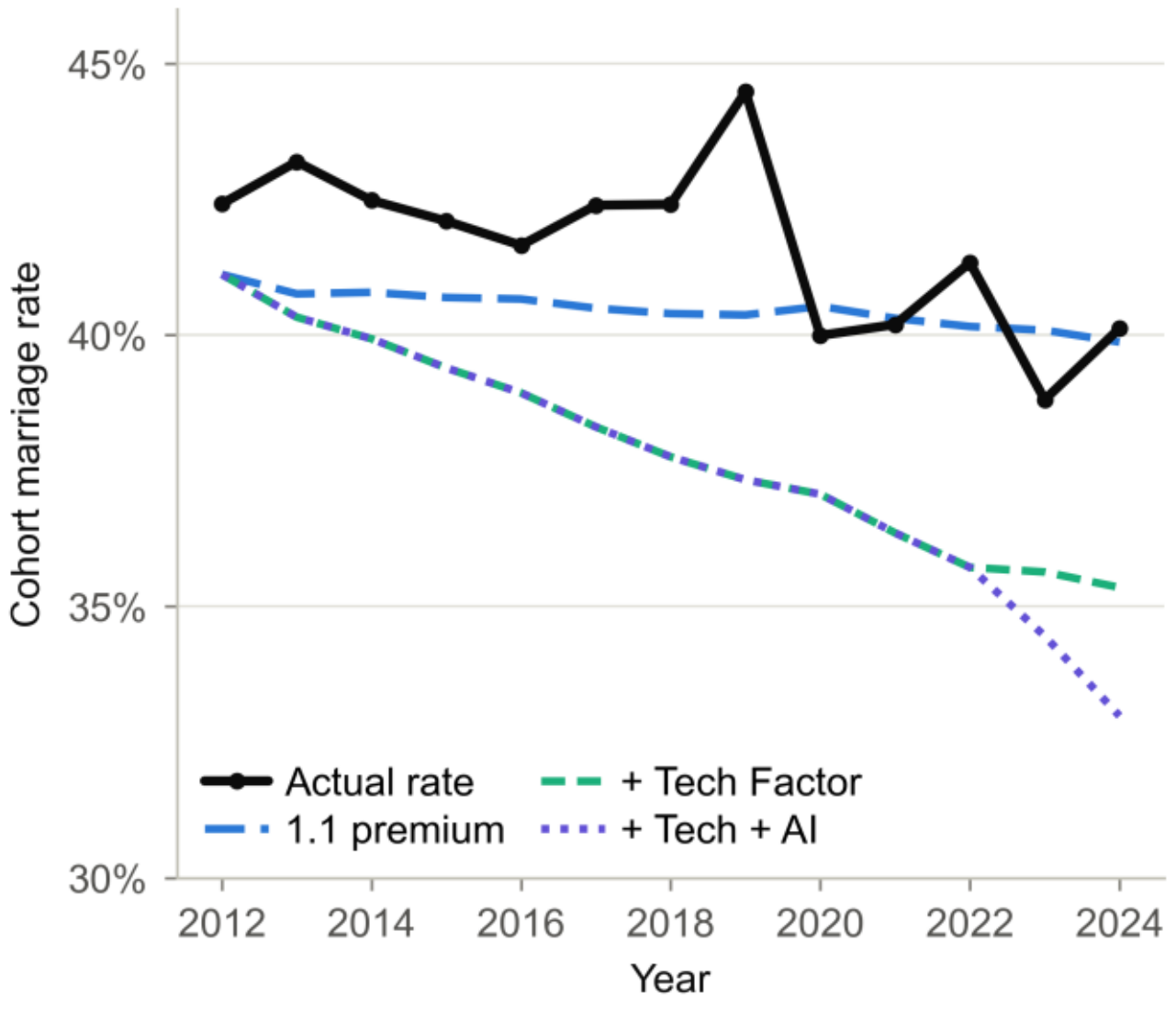

**Fig. S1** Cohort Marriage Rate Estimates (2012–2024): actual rates against the Fermi–Dirac model run at a baseline barrier of 1.1 times mean male salary, with the Tech Factor, and with the Tech and AI Factors. The rate supports the choice of those parameters

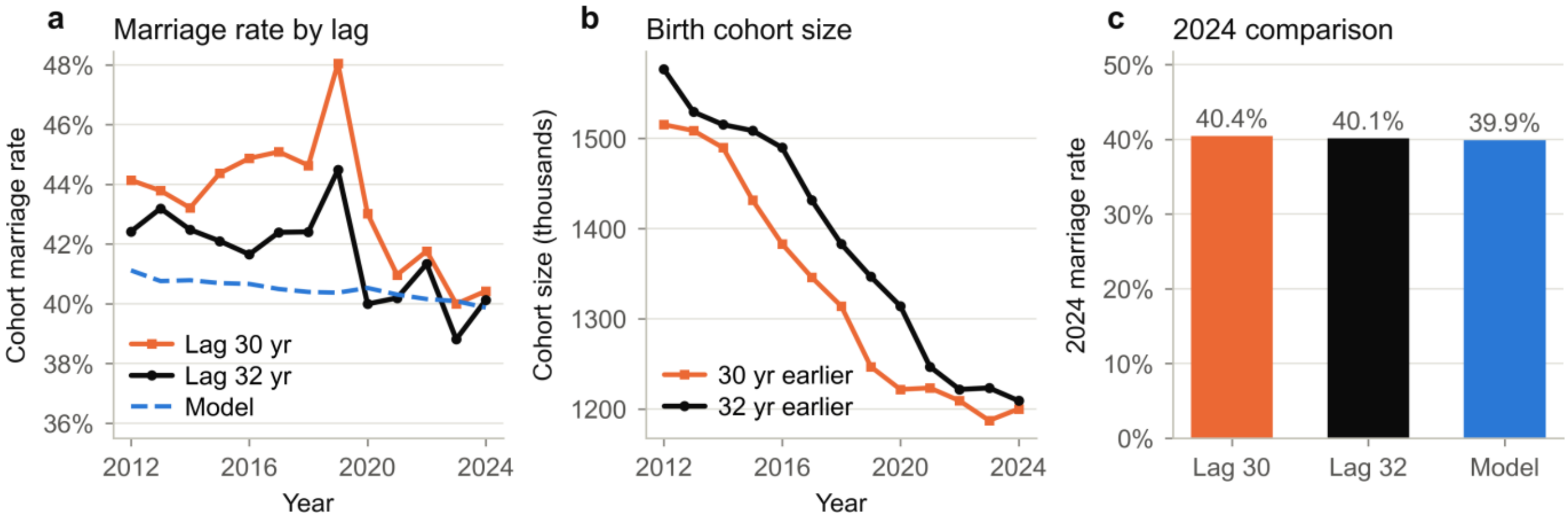

**Fig. S2** Effect of the birth-cohort lag on the observed marriage rate. (a) Rates calculated using births 30 and 32 years earlier, compared with the model. (b) Birth-cohort sizes used as the two denominators. (c) The 2024 rates under both lags and the model prediction

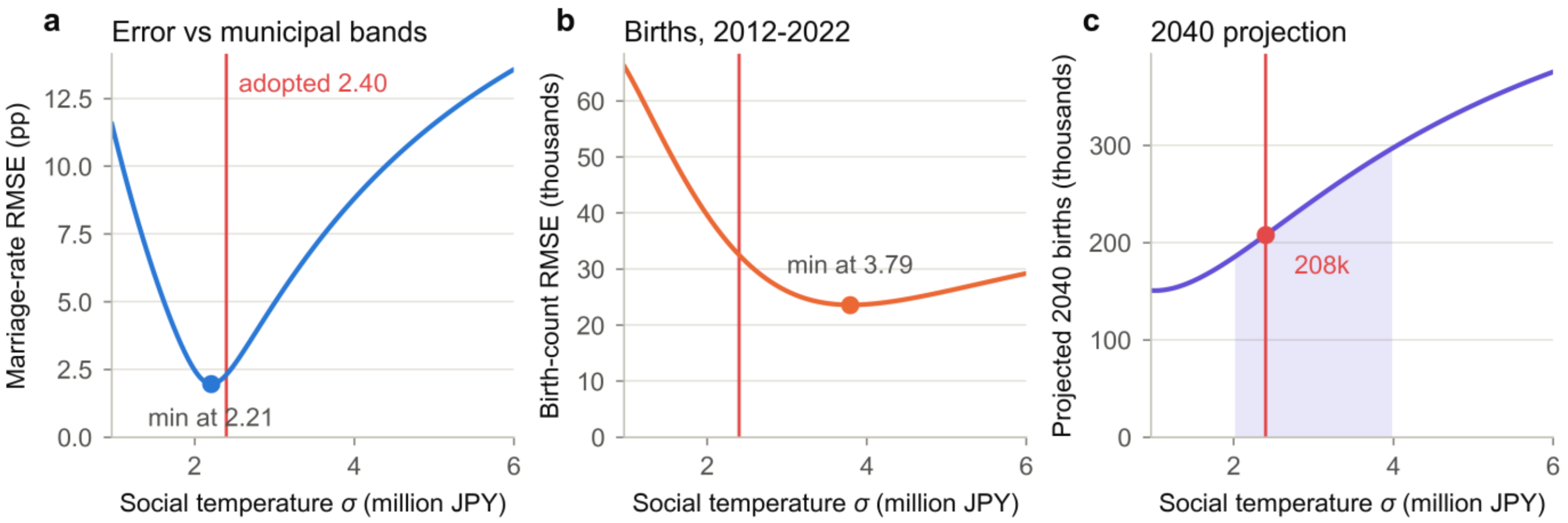

**Fig. S3** Sensitivity to social temperature σ, varied from 0.4 to 2.5 times the adopted value. (a) Marriage-rate RMSE against the University of Tokyo municipal income bands. (b) The root-mean-square error in predicted births over 2012–2022, with a broad minimum at a larger σ. The minimum-error values differ between the two comparisons; the red line marks the adopted σ = 2.40. (c) Projected 2040 births as σ varies

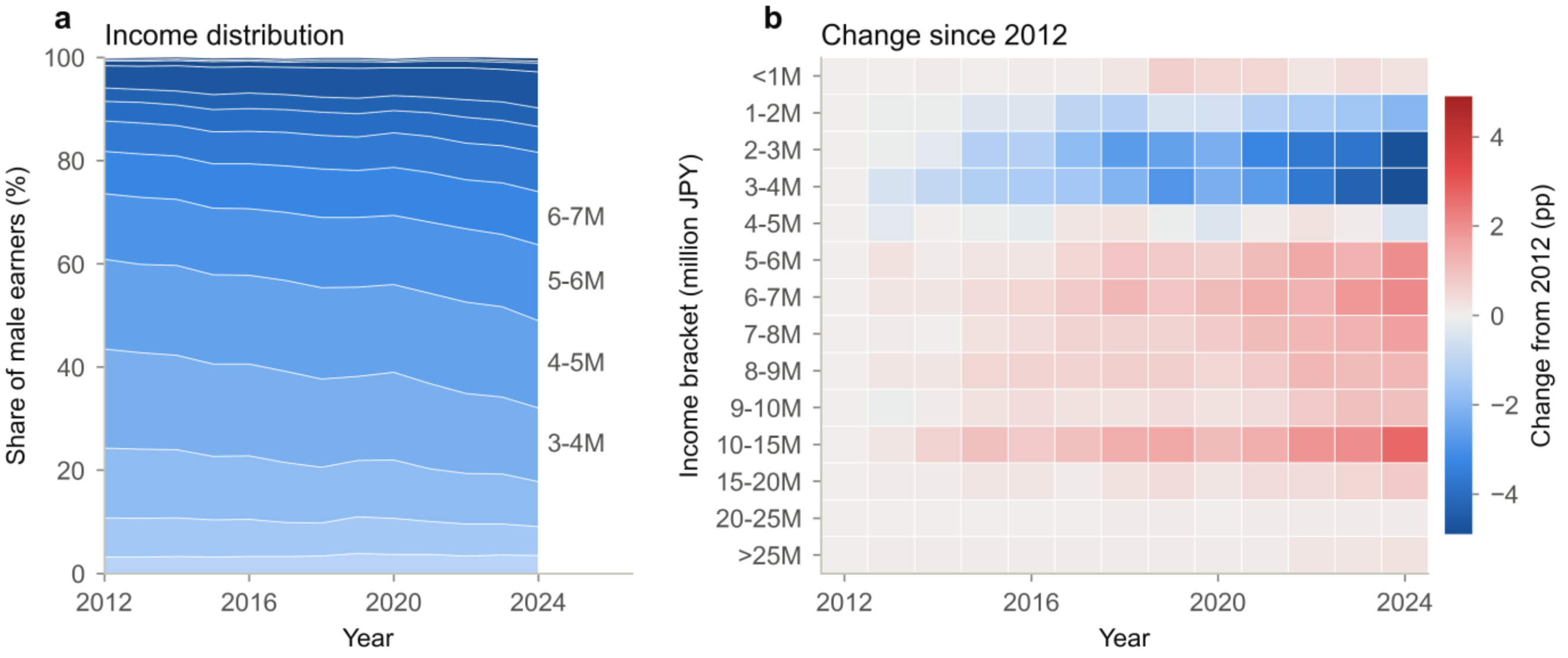

**Fig. S4** Male income distribution, 2012–2024. (a) Share of earners in each of the fourteen brackets published by the Ministry of Health, Labour and Welfare. (b) Change in each bracket's share relative to 2012, in percentage points: the share of earners fell in the 2–4M brackets and rose above 6M. M denotes million JPY

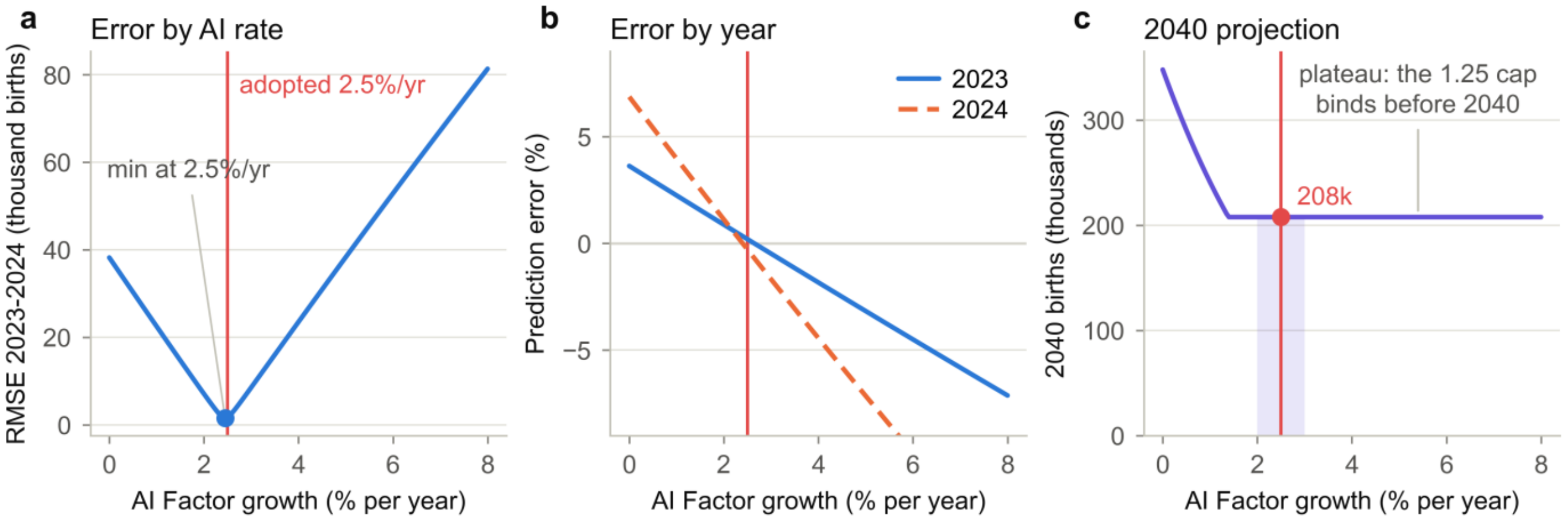

**Fig. S5** Calibration of the AI Factor growth rate, varied from zero to 3.2 times the adopted value. (a) The root-mean-square error in predicted births for 2023–2024. (b) Prediction errors for the two years shown separately. The red line marks the adopted increment of 0.025 per year. (c) Above roughly 1.4% per year, projected 2040 births remain unchanged because the AI Factor reaches its cap of 1.25 by 2040

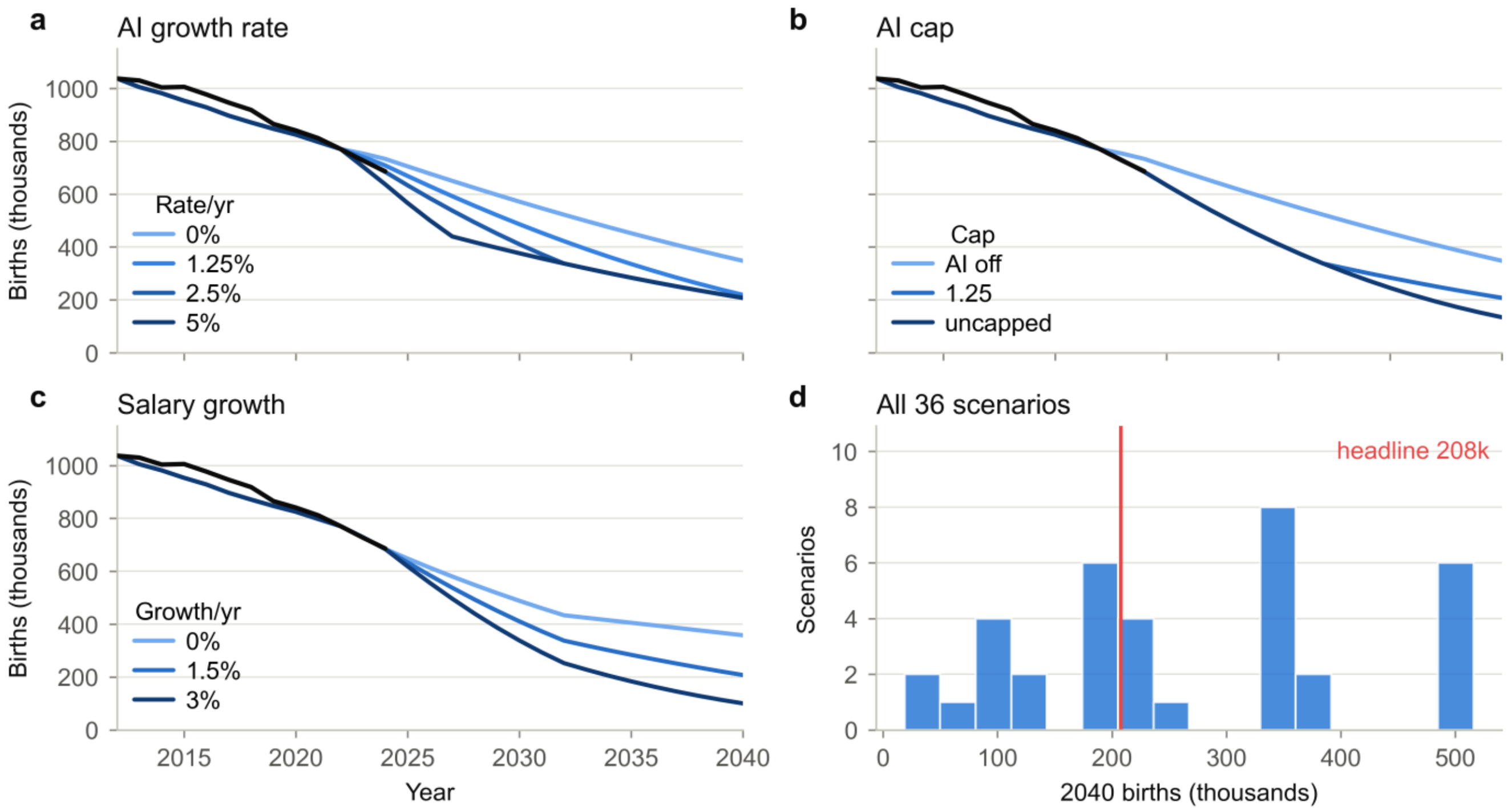

**Fig. S6** Birth projections under alternative model settings. (a) AI Factor growth at zero, half, the adopted rate and twice that rate. (b) AI Factor caps of 1.00 (no AI increase), 1.25 (adopted) and no cap. (c) Mean salary growth at zero, the adopted rate and twice that rate. Black lines in (a–c) show observed births. (d) Projected 2040 births across all 36 combinations; the red line marks the main projection of approximately 208,000

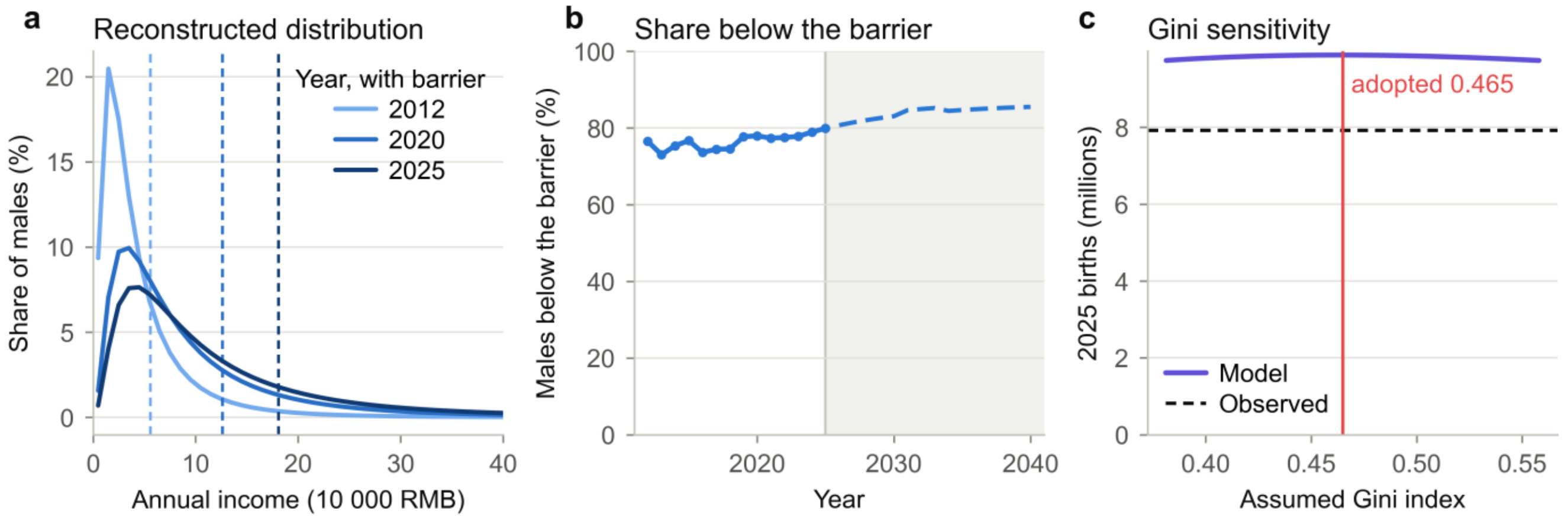

**Fig. S7** Income reconstruction for China. (a) Log-normal income distributions for 2012, 2020 and 2025; dashed vertical lines mark the corresponding marriage barriers. (b) Share of males below the barrier, with projections shown as a dashed line. (c) Predicted 2025 births as the assumed Gini index of income inequality varies from 0.82 to 1.20 times the adopted value of 0.465

**Table S1** Japan, TFR 1.0 after 2040. Every year from 2040 to 2100. The 2040 birth count is taken from Table 3. Births, women aged 20–39 and the age groups are in millions. The last column is the total fertility rate that would be needed in that year to reach the 2024 birth number

| Year | Births (M) | Women 20–39 (M) | <18 (M) | 18–64 (M) | 65+ (M) | Total (M) | Dependants per worker | TFR needed |
|---|---|---|---|---|---|---|---|---|
| 2040 | 0.208 | 10.10 | 7.32 | 56.81 | 33.92 | 98.05 | 0.73 | 1.36 |
| 2041 | 0.496 | 9.93 | 7.09 | 55.71 | 34.18 | 96.98 | 0.74 | 1.38 |
| 2042 | 0.487 | 9.74 | 6.89 | 54.64 | 34.28 | 95.81 | 0.75 | 1.41 |
| 2043 | 0.477 | 9.55 | 6.74 | 53.56 | 34.37 | 94.67 | 0.77 | 1.44 |
| 2044 | 0.467 | 9.34 | 6.62 | 52.50 | 34.40 | 93.53 | 0.78 | 1.47 |
| 2045 | 0.457 | 9.13 | 6.54 | 51.46 | 34.39 | 92.39 | 0.80 | 1.50 |
| 2046 | 0.444 | 8.88 | 6.49 | 50.43 | 34.30 | 91.22 | 0.81 | 1.54 |
| 2047 | 0.431 | 8.61 | 6.47 | 49.36 | 34.16 | 89.99 | 0.82 | 1.59 |
| 2048 | 0.416 | 8.32 | 6.48 | 48.27 | 33.95 | 88.69 | 0.84 | 1.65 |
| 2049 | 0.401 | 8.02 | 6.50 | 47.15 | 33.61 | 87.27 | 0.85 | 1.71 |
| 2050 | 0.385 | 7.70 | 6.55 | 46.06 | 33.69 | 86.29 | 0.87 | 1.78 |
| 2051 | 0.368 | 7.37 | 6.60 | 44.99 | 33.13 | 84.72 | 0.88 | 1.86 |
| 2052 | 0.351 | 7.03 | 6.65 | 43.95 | 32.61 | 83.20 | 0.89 | 1.95 |
| 2053 | 0.334 | 6.68 | 6.70 | 42.92 | 32.03 | 81.65 | 0.90 | 2.05 |
| 2054 | 0.317 | 6.34 | 6.75 | 41.94 | 31.34 | 80.03 | 0.91 | 2.17 |
| 2055 | 0.299 | 5.98 | 6.80 | 40.97 | 30.56 | 78.33 | 0.91 | 2.29 |
| 2056 | 0.282 | 5.64 | 6.84 | 39.98 | 29.75 | 76.57 | 0.92 | 2.43 |
| 2057 | 0.265 | 5.30 | 6.89 | 38.99 | 28.87 | 74.74 | 0.92 | 2.59 |
| 2058 | 0.248 | 4.97 | 6.93 | 38.01 | 28.02 | 72.96 | 0.92 | 2.76 |
| 2059 | 0.233 | 4.65 | 6.66 | 37.27 | 27.36 | 71.29 | 0.91 | 2.95 |
| 2060 | 0.217 | 4.34 | 6.39 | 36.57 | 26.72 | 69.68 | 0.91 | 3.16 |
| 2061 | 0.210 | 4.19 | 6.12 | 35.84 | 26.17 | 68.13 | 0.90 | 3.27 |
| 2062 | 0.203 | 4.05 | 5.86 | 35.12 | 25.65 | 66.63 | 0.90 | 3.39 |
| 2063 | 0.197 | 3.93 | 5.60 | 34.37 | 25.21 | 65.18 | 0.90 | 3.49 |
| 2064 | 0.191 | 3.82 | 5.35 | 33.64 | 24.81 | 63.80 | 0.90 | 3.59 |
| 2065 | 0.187 | 3.74 | 5.10 | 32.88 | 24.47 | 62.45 | 0.90 | 3.67 |
| 2066 | 0.183 | 3.67 | 4.87 | 32.12 | 24.13 | 61.12 | 0.90 | 3.74 |
| 2067 | 0.181 | 3.62 | 4.65 | 31.37 | 23.77 | 59.79 | 0.91 | 3.79 |
| 2068 | 0.179 | 3.58 | 4.44 | 30.63 | 23.41 | 58.48 | 0.91 | 3.83 |
| 2069 | 0.178 | 3.55 | 4.25 | 29.89 | 23.09 | 57.23 | 0.91 | 3.86 |
| 2070 | 0.177 | 3.54 | 4.08 | 29.18 | 22.77 | 56.02 | 0.92 | 3.87 |
| 2071 | 0.177 | 3.54 | 3.92 | 28.42 | 22.51 | 54.85 | 0.93 | 3.88 |
| 2072 | 0.177 | 3.55 | 3.78 | 27.64 | 22.29 | 53.72 | 0.94 | 3.87 |
| 2073 | 0.178 | 3.55 | 3.66 | 26.85 | 22.13 | 52.65 | 0.96 | 3.86 |
| 2074 | 0.178 | 3.56 | 3.56 | 26.06 | 21.98 | 51.60 | 0.98 | 3.85 |
| 2075 | 0.178 | 3.57 | 3.47 | 25.26 | 21.83 | 50.56 | 1.00 | 3.85 |
| 2076 | 0.179 | 3.58 | 3.40 | 24.46 | 21.67 | 49.53 | 1.03 | 3.84 |
| 2077 | 0.179 | 3.58 | 3.35 | 23.65 | 21.52 | 48.52 | 1.05 | 3.83 |
| 2078 | 0.179 | 3.59 | 3.31 | 22.84 | 21.31 | 47.46 | 1.08 | 3.82 |
| 2079 | 0.180 | 3.59 | 3.28 | 22.04 | 21.13 | 46.45 | 1.11 | 3.82 |
| 2080 | 0.180 | 3.60 | 3.26 | 21.24 | 20.93 | 45.43 | 1.14 | 3.81 |
| 2081 | 0.173 | 3.46 | 3.23 | 20.46 | 20.71 | 44.41 | 1.17 | 3.97 |

| Year | Births (M) | Women 20–39 (M) | <18 (M) | 18–64 (M) | 65+ (M) | Total (M) | Dependants per worker | TFR needed |
|---|---|---|---|---|---|---|---|---|
| 2082 | 0.166 | 3.32 | 3.21 | 19.71 | 20.45 | 43.37 | 1.20 | 4.13 |
| 2083 | 0.159 | 3.18 | 3.18 | 18.98 | 20.20 | 42.35 | 1.23 | 4.31 |
| 2084 | 0.152 | 3.05 | 3.15 | 18.29 | 19.87 | 41.31 | 1.26 | 4.50 |
| 2085 | 0.146 | 2.92 | 3.12 | 17.63 | 19.54 | 40.29 | 1.28 | 4.71 |
| 2086 | 0.139 | 2.79 | 3.08 | 17.00 | 19.20 | 39.27 | 1.31 | 4.92 |
| 2087 | 0.133 | 2.67 | 3.03 | 16.41 | 18.85 | 38.28 | 1.33 | 5.15 |
| 2088 | 0.128 | 2.55 | 2.98 | 15.86 | 18.46 | 37.30 | 1.35 | 5.38 |
| 2089 | 0.122 | 2.44 | 2.93 | 15.35 | 18.09 | 36.36 | 1.37 | 5.62 |
| 2090 | 0.117 | 2.34 | 2.87 | 14.89 | 17.63 | 35.39 | 1.38 | 5.86 |
| 2091 | 0.112 | 2.25 | 2.80 | 14.49 | 17.12 | 34.41 | 1.38 | 6.11 |
| 2092 | 0.108 | 2.16 | 2.73 | 14.13 | 16.57 | 33.42 | 1.37 | 6.34 |
| 2093 | 0.104 | 2.09 | 2.66 | 13.81 | 15.99 | 32.46 | 1.35 | 6.58 |
| 2094 | 0.101 | 2.02 | 2.58 | 13.54 | 15.37 | 31.49 | 1.33 | 6.80 |
| 2095 | 0.098 | 1.96 | 2.50 | 13.31 | 14.73 | 30.54 | 1.29 | 7.00 |
| 2096 | 0.095 | 1.91 | 2.41 | 13.11 | 14.07 | 29.59 | 1.26 | 7.19 |
| 2097 | 0.093 | 1.87 | 2.33 | 12.96 | 13.37 | 28.66 | 1.21 | 7.35 |
| 2098 | 0.092 | 1.83 | 2.24 | 12.82 | 12.69 | 27.75 | 1.16 | 7.48 |
| 2099 | 0.090 | 1.81 | 2.16 | 12.69 | 11.98 | 26.83 | 1.11 | 7.59 |
| 2100 | 0.090 | 1.79 | 2.08 | 12.57 | 11.29 | 25.94 | 1.06 | 7.67 |

**Table S2** China, TFR 1.0 after 2040. Every year from 2040 to 2100. The 2040 birth count is taken from Table 5. Births, women aged 20–39 and the age groups are in millions. The last column is the total fertility rate that would be needed in that year to reach the 2024 birth number

| Year | Births (M) | Women 20–39 (M) | <18 (M) | 18–64 (M) | 65+ (M) | Total (M) | Dependants per worker | TFR needed |
|---|---|---|---|---|---|---|---|---|
| 2040 | 4.591 | 139.74 | 121.30 | 733.52 | 324.44 | 1179.26 | 0.61 | 1.37 |
| 2041 | 6.937 | 138.74 | 119.22 | 724.93 | 328.03 | 1172.18 | 0.62 | 1.38 |
| 2042 | 6.856 | 137.12 | 116.54 | 716.64 | 317.99 | 1151.17 | 0.61 | 1.39 |
| 2043 | 6.770 | 135.40 | 115.39 | 705.73 | 312.54 | 1133.66 | 0.61 | 1.41 |
| 2044 | 6.683 | 133.65 | 112.84 | 696.01 | 306.69 | 1115.54 | 0.60 | 1.43 |
| 2045 | 6.550 | 130.99 | 110.76 | 688.70 | 298.44 | 1097.90 | 0.59 | 1.46 |
| 2046 | 6.440 | 128.80 | 109.15 | 680.51 | 292.90 | 1082.57 | 0.59 | 1.48 |
| 2047 | 6.308 | 126.16 | 107.95 | 667.50 | 287.35 | 1062.81 | 0.59 | 1.51 |
| 2048 | 6.152 | 123.04 | 107.10 | 655.67 | 281.17 | 1043.95 | 0.59 | 1.55 |
| 2049 | 5.983 | 119.66 | 106.56 | 644.11 | 273.75 | 1024.42 | 0.59 | 1.59 |
| 2050 | 5.799 | 115.99 | 106.28 | 631.56 | 268.45 | 1006.29 | 0.59 | 1.64 |
| 2051 | 5.611 | 112.22 | 106.01 | 618.31 | 264.39 | 988.70 | 0.60 | 1.70 |
| 2052 | 5.367 | 107.35 | 105.69 | 603.33 | 262.18 | 971.20 | 0.61 | 1.78 |
| 2053 | 5.118 | 102.36 | 105.31 | 588.77 | 260.90 | 954.98 | 0.62 | 1.86 |
| 2054 | 4.853 | 97.06 | 104.85 | 574.43 | 260.97 | 940.25 | 0.64 | 1.97 |
| 2055 | 4.591 | 91.82 | 104.32 | 559.47 | 263.44 | 927.23 | 0.66 | 2.08 |
| 2056 | 4.294 | 85.87 | 103.67 | 547.39 | 262.63 | 913.70 | 0.67 | 2.22 |
| 2057 | 4.007 | 80.13 | 102.91 | 536.20 | 259.76 | 898.87 | 0.68 | 2.38 |
| 2058 | 3.763 | 75.26 | 102.08 | 525.87 | 255.73 | 883.68 | 0.68 | 2.54 |
| 2059 | 3.529 | 70.57 | 98.67 | 518.83 | 253.77 | 871.27 | 0.68 | 2.70 |
| 2060 | 3.353 | 67.06 | 95.17 | 512.13 | 251.10 | 858.40 | 0.68 | 2.85 |
| 2061 | 3.266 | 65.32 | 91.67 | 505.58 | 243.90 | 841.14 | 0.66 | 2.92 |
| 2062 | 3.202 | 64.04 | 88.19 | 499.41 | 237.92 | 825.51 | 0.65 | 2.98 |
| 2063 | 3.148 | 62.97 | 84.78 | 493.95 | 231.84 | 810.57 | 0.64 | 3.03 |
| 2064 | 3.081 | 61.61 | 81.43 | 488.90 | 224.70 | 795.02 | 0.63 | 3.10 |
| 2065 | 3.048 | 60.97 | 78.17 | 482.84 | 217.93 | 778.93 | 0.61 | 3.13 |
| 2066 | 2.982 | 59.64 | 75.00 | 476.26 | 210.00 | 761.25 | 0.60 | 3.20 |
| 2067 | 2.927 | 58.54 | 71.94 | 469.27 | 202.91 | 744.12 | 0.59 | 3.26 |
| 2068 | 2.882 | 57.65 | 69.02 | 462.42 | 195.91 | 727.35 | 0.57 | 3.31 |
| 2069 | 2.846 | 56.92 | 66.26 | 454.80 | 189.06 | 710.12 | 0.56 | 3.35 |
| 2070 | 2.818 | 56.35 | 63.71 | 446.64 | 185.57 | 695.92 | 0.56 | 3.39 |
| 2071 | 2.796 | 55.92 | 61.39 | 437.90 | 183.47 | 682.75 | 0.56 | 3.41 |
| 2072 | 2.779 | 55.58 | 59.31 | 428.56 | 182.74 | 670.61 | 0.56 | 3.43 |
| 2073 | 2.761 | 55.22 | 57.48 | 418.52 | 183.39 | 659.39 | 0.58 | 3.46 |
| 2074 | 2.741 | 54.82 | 55.93 | 408.17 | 184.48 | 648.57 | 0.59 | 3.48 |
| 2075 | 2.720 | 54.39 | 54.64 | 397.43 | 185.90 | 637.97 | 0.61 | 3.51 |
| 2076 | 2.696 | 53.91 | 53.57 | 386.70 | 187.54 | 627.82 | 0.62 | 3.54 |
| 2077 | 2.669 | 53.38 | 52.71 | 373.88 | 191.88 | 618.48 | 0.65 | 3.57 |
| 2078 | 2.641 | 52.82 | 52.00 | 360.84 | 196.79 | 609.63 | 0.69 | 3.61 |
| 2079 | 2.612 | 52.24 | 51.35 | 347.23 | 201.29 | 599.87 | 0.73 | 3.65 |
| 2080 | 2.582 | 51.65 | 50.73 | 333.88 | 205.11 | 589.72 | 0.77 | 3.69 |
| 2081 | 2.495 | 49.91 | 50.08 | 319.17 | 210.00 | 579.25 | 0.81 | 3.82 |

| Year | Births (M) | Women 20–39 (M) | <18 (M) | 18–64 (M) | 65+ (M) | Total (M) | Dependants per worker | TFR needed |
|---|---|---|---|---|---|---|---|---|
| 2082 | 2.409 | 48.18 | 49.40 | 305.02 | 214.58 | 569.01 | 0.87 | 3.96 |
| 2083 | 2.323 | 46.46 | 48.68 | 292.84 | 216.58 | 558.10 | 0.91 | 4.11 |
| 2084 | 2.238 | 44.75 | 47.93 | 281.17 | 217.70 | 546.81 | 0.94 | 4.26 |
| 2085 | 2.155 | 43.09 | 47.16 | 272.10 | 215.84 | 535.10 | 0.97 | 4.43 |
| 2086 | 2.073 | 41.45 | 46.35 | 264.36 | 212.27 | 522.99 | 0.98 | 4.60 |
| 2087 | 1.993 | 39.85 | 45.50 | 257.65 | 207.20 | 510.35 | 0.98 | 4.79 |
| 2088 | 1.915 | 38.30 | 44.60 | 251.45 | 201.57 | 497.61 | 0.98 | 4.98 |
| 2089 | 1.841 | 36.81 | 43.64 | 244.70 | 196.37 | 484.71 | 0.98 | 5.18 |
| 2090 | 1.770 | 35.40 | 42.63 | 239.56 | 189.80 | 471.99 | 0.97 | 5.39 |
| 2091 | 1.703 | 34.07 | 41.57 | 233.09 | 182.68 | 457.35 | 0.96 | 5.60 |
| 2092 | 1.642 | 32.84 | 40.48 | 227.21 | 174.91 | 442.59 | 0.95 | 5.81 |
| 2093 | 1.586 | 31.72 | 39.34 | 221.88 | 166.08 | 427.31 | 0.93 | 6.01 |
| 2094 | 1.536 | 30.72 | 38.18 | 217.07 | 157.04 | 412.29 | 0.90 | 6.21 |
| 2095 | 1.492 | 29.83 | 37.01 | 212.73 | 146.18 | 395.92 | 0.86 | 6.40 |
| 2096 | 1.454 | 29.08 | 35.82 | 208.85 | 135.48 | 380.15 | 0.82 | 6.56 |
| 2097 | 1.422 | 28.44 | 34.63 | 205.38 | 126.33 | 366.34 | 0.78 | 6.71 |
| 2098 | 1.396 | 27.91 | 33.44 | 202.08 | 117.57 | 353.09 | 0.75 | 6.84 |
| 2099 | 1.374 | 27.48 | 32.32 | 198.89 | 111.25 | 342.46 | 0.72 | 6.94 |
| 2100 | 1.356 | 27.11 | 31.27 | 195.80 | 106.13 | 333.19 | 0.70 | 7.04 |